\documentclass{article}
\usepackage{ourMacros}           
\usepackage{abbreviations}       
\usepackage[latin1]{rtthesis-captions}
\usepackage{rtthesis-example}
\usepackage{epstopdf}

\newcommand\apfw{\nu} 
\newcommand{\+}{\mathsf{T}}                      

\title{Sequential Monte Carlo methods for \\ system identification} 

\author{Thomas B. Sch\"on, Fredrik Lindsten, Johan Dahlin, \\ Johan W{\aa}gberg, Christian A. Naesseth, \\ Andreas Svensson and Liang Dai%
\thanks{This work was supported by the projects \emph{Learning of complex dynamical
    systems} (Contract number: 637-2014-466) and \emph{Probabilistic modeling of dynamical systems}
  (Contract number: 621-2013-5524), both funded by the Swedish Research Council. TS, JW, AS and LD are with Division of Systems and Control, Uppsala University, Uppsala, Sweden. E-mail: \texttt{\{thomas.schon, johan.wagberg, andreas.svensson, liang.dai\}@it.uu.se}. FL is with the Department of Engineering, University of Cambridge, Cambridge, United Kingdom. E-mail: \texttt{fredrik.lindsten@eng.cam.ac.uk}. JD and CAN are with the Division of Automatic Control, Link{\"o}ping University, Link{\"o}ping, Sweden. E-mail: \texttt{ \{christian.a.naesseth,johan.dahlin\}@liu.se}.  }%
}

\begin{document}

\newcommand{\coverTitle}{Sequential Monte Carlo methods for system identification}
\newcommand{\coverAuthors}{Thomas B. Sch\"on, Fredrik Lindsten, Johan Dahlin, Johan W{\aa}gberg, Christian A. Naesseth, Andreas Svensson and Liang Dai}
\newcommand{\coverYear}{2015}
\newcommand{\coverStatus}{Accepted for publication.}

\begin{titlepage}
\begin{center}
{\large \em Technical report}

\vspace*{2.5cm}
%
{\Huge \bfseries \coverTitle  \\[0.4cm]}

%
{\Large \coverAuthors \\[2cm]}

\renewcommand\labelitemi{\color{red}\large$\bullet$}
\begin{itemize}
\item {\Large \textbf{Please cite this version:}} \\[0.4cm]
\large
\coverAuthors. \coverTitle. In \textit{Proceedings of the 17th IFAC
  Symposium on System Identification  (SYSID)},
Beijing, China, October 2015.

\end{itemize}

\vfill

\begin{abstract}
  One of the key challenges in identifying nonlinear and possibly
  non-Gaussian state space models (\ssm{s}) is the intractability of
  estimating the system state. Sequential Monte Carlo (\smc) methods,
  such as the particle filter (introduced more than two decades ago),
  provide numerical solutions to the nonlinear state estimation
  problems arising in \ssm{s}. When combined with additional
  identification techniques, these algorithms provide solid solutions
  to the nonlinear system identification problem. We describe two
  general strategies for creating such combinations and discuss why
  \smc is a natural tool for implementing these strategies.
\end{abstract}

\vfill
\end{center}
\end{titlepage}

\maketitle
\thispagestyle{empty}
\pagestyle{empty}

\begin{abstract}
  One of the key challenges in identifying
  nonlinear and possibly non-Gaussian state space models (\ssm{s}) is the intractability of
  estimating the system state. Sequential Monte Carlo (\smc) methods, such as the particle filter
  (introduced more than two decades ago), provide numerical solutions to the nonlinear state
  estimation problems arising in \ssm{s}. When combined with additional identification techniques,
  these algorithms provide solid solutions to the nonlinear system identification problem. We
  describe two general strategies for creating such combinations and discuss why \smc is a natural
  tool for implementing these strategies.
\end{abstract}

\clearpage

\section{Introduction}
This paper is concerned with system identification of nonlinear state space models (\ssm{s})
in discrete time. The general model that we consider is given by,
\begin{subequations}
  \label{eq:SSM}
  \begin{align}
    \label{eq:SSMa}
    x_{t+1}\mid x_t &\sim \f_{\theta}(x_{t+1}\mid x_t, u_t),\\
    \label{eq:SSMb}
    y_{t}\mid x_t &\sim \g_{\theta}(y_{t}\mid x_t, u_t),\\
    \label{eq:SSMc}
    (\theta &\sim  \prior(\theta)),
  \end{align}
\end{subequations}
where the states, the known inputs and the observed measurements are denoted by $x_t\in \setX
\subseteq \R^{\nX}$, $u_t\in \setU \subseteq \R^{\nU}$ and $y_t\in \setY \subseteq \R^{\nY}$,
respectively. The dynamics and the measurements are modeled by the probability density functions
(\pdf{s}) $\f_{\theta}(\cdot)$ and $\g_{\theta}(\cdot)$, respectively, parameterised by the unknown
vector $\theta \in \setTh \subseteq \R^{\nTh}$. The initial state~$x_1$ is distributed according to
some distribution $\mu_{\theta}(x_1)$.
For notational simplicity, we will from hereon (without loss of generality)
drop the known input $u_t$ from the notation.
When considering Bayesian identification, meaning that~$\theta$ is modelled as an unobserved random
variable, we also place a prior $\prior(\theta)$ on~$\theta$.  We are concerned with off-line
identification, \ie we wish to find the unknown parameters~$\theta$ in~\eqref{eq:SSM} based on a
\emph{batch} of~$\T$ measurements.

The \emph{key challenge} that will drive the development throughout this paper is how to deal with
the difficulty that the states $x_{1:\T}$ in~\eqref{eq:SSM} are unknown. 
We will distinguish between two different strategies for handling this:
\begin{enumerate}
\item \textbf{Marginalisation } amounts to marginalising (integrating out) the states from
  the problem and viewing $\theta$ as the only unknown quantity of interest. In the frequentistic
  problem formulation, the prediction error method \citep{Ljung:1999} and direct maximisation of the
  likelihood belong to this strategy. In the Bayesian formulation, the Metropolis--Hastings algorithm
  \citep{Metropolis:1953,Hastings:1970} can be used to approximate the posterior distribution of the parameters
  conditionally on the data.
\item \textbf{Data augmentation } treats the states as auxiliary variables that are estimated
  together with the parameters. The expectation maximisation (EM) algorithm of \citet{DempsterLR:1977}
  solves the maximum likelihood formulation in this way and the Gibbs sampler of~\citet{GemanG:1984}
  solves the Bayesian problem by this strategy.
\end{enumerate}
In the special case when the model \eqref{eq:SSM} is linear and Gaussian, there are closed form expressions
available from the Kalman filter and the associated smoother. The primary focus of this
paper, however, is the more challenging nonlinear and/or non-Gaussian case.
More than two decades ago \citep[\eg,][]{Gordon:1993,Kitagawa:1993} \emph{sequential Monte Carlo} (\smc) methods started to
emerge with the introduction of the \emph{particle filter}. These methods have since then undergone a 
rapid development and today they constitute a standard solution to the problem of computing
the latent (\ie unobserved/unknown/hidden) states in nonlinear/non-Gaussian \ssm{s}.

%
The \emph{aim} of this paper is to show how \smc can be used in solving the nonlinear system
identification problems that arise in finding the unknown parameters in~\eqref{eq:SSM}. We do not
aim to cover all different methods that are available, but instead we aim to clearly describe
exactly where and how the need for \smc arises and focus on the key principles. Complementary
overview paper are provided by \citet{KantasDSMC:2014} and by \citet{AndrieuDST:2004}.

%
We consider the Bayesian and the maximum likelihood formulations, as defined in
Section~\ref{sec:pf}. The rest of the paper is divided into three parts. In the first part
(Sections~\ref{sec:IdStrat:Marg}~and~\ref{sec:IdStrat:DA}) we describe the marginalisation and data
augmentation strategies and show where the need for \smc arise. The second part
(Section~\ref{sec:SMC}) provides a rather self-contained derivation of the particle filter and
outlines some of its properties. Finally, in the third part
(Section~\ref{sec:NL:marg}--\ref{sec:IdMarkovKernel}) we show how these \emph{particle methods} can
be used to implement the identification strategies described in the first part, resulting in several
state-of-the-art algorithms for nonlinear system identification. Loosely speaking, the \smc-part of
the various algorithms that we introduce is essentially a way of systematically exploring the state
space $\setX^{\T}$ in a nonlinear \ssm~\eqref{eq:SSM} in order to address the key challenge of
dealing with the latent state sequence $x_{1:\T}$.


\section{Problem formulation}
\label{sec:pf}
%
There are different ways in which the system identification problem can be formulated. Two
common formalisms are grounded in frequentistic and Bayesian statistics, respectively.
We will treat both of these formulations in this paper, without making any
individual ranking among them. 
First, we consider the frequentistic, or \emph{maximum likelihood} (\ml), formulation. This amounts
to finding a point estimate of the unknown parameter $\theta$, for which the observed data is as likely
as possible. This is done by maximising the data likelihood function according to
\begin{align}
  \label{eq:ML}
  \widehat{\theta}_{\text{ML}} = \argmax{\theta\in\setTh}{p_{\theta}(y_{1:\T})}
  = \argmax{\theta\in\setTh}{\log p_{\theta}(y_{1:\T})}.
\end{align}
For a thorough treatment of the use of \ml for system identification, see \eg,
\citet{Ljung:1999,SoderstromS:1989}. 

Secondly, in the \emph{Bayesian} formulation, the unknown parameters $\theta$ are modeled as a
random variable (or random vector) according to~\eqref{eq:SSMc}. The system identification problem
thus amounts to computing the \emph{posterior} distribution of $\theta$ given the observed
data. According to Bayes' theorem, the posterior distribution is given by,
\begin{align}
  \label{eq:Bayes}
  p(\theta\mid y_{1:\T}) = \frac{p_{\theta}(y_{1:\T})\prior(\theta)}{p(y_{1:\T})}.
\end{align}
Note that the likelihood function should now be interpreted as the conditional \pdf of the
observations given the parameters $\theta$, \ie $p_{\theta}(y_{1:\T}) = p(y_{1:\T} \mid \theta)$.
However, to be able to discuss the different identification criteria in a common setting, we will,
with a slight abuse of notation, denote the likelihood by $p_{\theta}(y_{1:\T})$ also in the
Bayesian formulation. An early account of Bayesian system identification is provided by
\citet{Peterka:1981} and a more recent description is available in \citet{NinnessH:2010}.

The \emph{central object} in both formulations above is the observed data likelihood
$p_{\theta}(y_{1:\T})$, or its constituents $p_{\theta}(y_t\mid y_{1:t-1})$ as,
\begin{align}
  \label{eq:likelihood}
  p_{\theta}(y_{1:\T}) = \prod_{t=1}^{\T} p_{\theta}(y_t\mid y_{1:t-1}),
\end{align}
where we have used the convention $y_{1\mid 0} \triangleq \emptyset$. For the nonlinear
\ssm~\eqref{eq:SSM}, the likelihood~\eqref{eq:likelihood} is not available in closed form. This is a
result of the fact that the latent state sequence~$x_{1:\T}$, \ie $p(x_{1:\T}\mid y_{1:\T})$, is
unknown. Indeed, a relationship between the likelihood and the latent states can be obtained via
marginalization of the joint density $p_{\theta}(x_{1:\T}, y_{1:\T})$ \wrt $x_{1:\T}$ according to,
\begin{align}
  \label{eq:marg}
  p_{\theta}(y_{1:\T}) = \int p_{\theta}(x_{1:\T}, y_{1:\T}) \myd x_{1:\T},
\end{align}
where the model provides a closed form expression for the integrand according to,
\begin{align}
  \label{eq:jointXY}
  p_{\theta}(x_{1:\T}, y_{1:\T}) = \mu_\theta(x_1)\prod_{t=1}^{\T}\g_{\theta}(y_t\mid x_t) \prod_{t=1}^{\T-1}\f_{\theta}(x_{t+1}\mid x_t).
\end{align}
This expression does not involve any marginalisation, whereas the observed data likelihood
$p_{\theta}(y_{1:\T})$ is found by averaging the joint distribution $p_{\theta}(x_{1:\T}, y_{1:\T})$
over all possible state sequences according to~\eqref{eq:marg}. Equivalently, we can express
$p_{\theta}(y_{1:\T})$ as in~\eqref{eq:likelihood} where the one-step predictive likelihood can be
written (using marginalisation) as,
\begin{align}
  \label{eq:marg2}
  p_{\theta}(y_t\mid y_{1:t-1}) = \int \g_{\theta}(y_t\mid x_t) p_{\theta}(x_t\mid y_{1:t-1})\myd x_t.
\end{align}
These expressions highlight the tight relationship between the system identification problem and the
state inference problem. A key challenge that will drive the developments in this work is how
to deal with the latent states. For nonlinear system identification, the need for 
computational methods, such as \smc, is tightly coupled to the intractability of the integrals
in~\eqref{eq:marg}~and~\eqref{eq:marg2}.

To illustrate the strategies and algorithms introduced we will use them to solve two concrete
problems, which are formulated below.

\begin{example}[Linear Gaussian model]
  \label{example:LGSS}%
  %
  Our first illustrative example is a simple linear Gaussian state space (\lgss) model, given by
  \begin{subequations}
    \label{eq:LGSS}
    \begin{align}
      \label{eq:LGSSa}
      x_{t+1} &= 0.7x_t + v_t, \quad &v_t& \sim \N(0,\theta^{-1}),\\
      \label{eq:LGSSb}
      y_t &= 0.5x_t + e_t, \quad &e_t& \sim \N(0,0.1),\\
      \label{eq:LGSSc}
      (\theta &\sim \gam(0.01,0.01) ),
    \end{align}
  \end{subequations}
  where the unknown parameter $\theta$ corresponds to the \emph{precision} (inverse variance) of the
  process noise $v_t$. Note that the prior for the Bayesian model is chosen as the Gamma ($\gam$)
  distribution with known parameters, for reasons of simplicity, since this is the \emph{conjugate
    prior} for this model. The initial distribution $\mu_\theta(x_1)$ is chosen as the stationary
  distribution of the state process. Identification of $\theta$ is based on a simulated data set
  consisting of $T = 100$ samples $y_{1:100}$ with true parameter $\theta_0 = 1$. 
\end{example}

\begin{example}[Nonlinear non-Gaussian model]
  \label{example:Nonlinear}%
  Our second example, involving real data, is related to a problem in paleoclimatology.
  \citet{ShumwayS:2011} considered the problem of modelling the thickness of ice varves (layers of
  sediments that are deposited from melting glaciers). The silt and sand that is deposited over one
  year makes up one varve and changes in the varve thicknesses indicates temperature changes. The
  data set that is used contains the thickness of $634$ ice varves formed at a location in
  Massachusetts between years $9\thinspace883$ and $9\thinspace250$~BC.  
  We make use of a nonlinear and non-Gaussian \ssm proposed by \citet{Langrock:2011} to model this
  data:
  \begin{subequations}
    \label{eq:icevarvemodel}%
    \begin{align}
      x_{t+1} \mid x_t &\sim \mathcal{N}(x_{t+1}; \phi x_t,\tau^{-1}), \\
      y_{t} \mid x_t   &\sim \mathcal{G}(y_t; 6.25, 0.256 \exp(-x_t) ),
    \end{align}%
  \end{subequations}%
  \noindent with parameters $\theta=\{\phi,\tau\}$. 
  The initial distribution $\mu_\theta(x_1)$ is chosen as the stationary distribution of the state process.
  The data set and a more complete description of the problem is provided by \cite{ShumwayS:2011}.
\end{example}




\section{Identification strategy -- marginalisation}
\label{sec:IdStrat:Marg}
The marginalisation strategy amounts to solving the identification problem---either \eqref{eq:ML} or
\eqref{eq:Bayes}---by computing the integral appearing in~\eqref{eq:marg2} (or, equivalently,
\eqref{eq:marg}). That is, we \emph{marginalize} (integrate out) the latent states~$x_{1:T}$ and
view~$\theta$ as the only unknown quantity of interest.

In some special cases the marginalisation can be done exactly. In particular, for LGSS models the
one-step predictive density in \eqref{eq:marg2} is Gaussian and computable by running a Kalman
filter.  For the general case, however, some numerical approximation of the integral
in~\eqref{eq:marg2} is needed.  We will elaborate on this in Section~\ref{sec:NL:marg}, where we
investigate the possibility of using SMC to perform the marginalisation. For now, however, to
illustrate the general marginalisation strategy, we will assume that the integral
in~\eqref{eq:marg2} can be solved exactly.

\subsection{ML identification via direct optimisation}
%
Consider first the ML formulation~\eqref{eq:ML}. Direct optimisation (DO) amounts to working
directly on the problem
\begin{align}
  \label{eq:DO:ML}
  \widehat{\theta}_{\text{ML}} = \argmax{\theta\in\setTh}{ \sum_{t=1}^{\T} \log
    \int \g_{\theta}(y_t\mid x_t) p_{\theta}(x_t\mid y_{1:t-1})\myd x_{t}},
\end{align}
where we have rewritten the data log-likelihood using~\eqref{eq:likelihood} and~\eqref{eq:marg2}.
Even though we momentarily neglect the difficulty in computing the integral above
it is typically not possible to solve the optimisation problem \eqref{eq:DO:ML} in closed form.
Instead, we have to resort to numerical optimisation methods, see \eg
\cite{NocedalW:2006}. These methods typically find a maximiser of the log-likelihood function $\log
p_{\theta}(y_{1:\T})$ by iteratively refining an estimate $\iteratesPE{\theta}{k}$ of the maximiser
$\widehat{\theta}_{\text{ML}}$ according to
\begin{align}
  \label{eq:DO:iterates}
  \iteratesPE{\theta}{k+1} = \iteratesPE{\theta}{k} + \iteratesPE{\alpha}{k}\iteratesPE{s}{k}.
\end{align}
Here, $\iteratesPE{s}{k}$ denotes the search direction which is computed based on information about
the cost function available from previous iterations. The step size $\iteratesPE{\alpha}{k}$, tells
us how far we should go in the search direction. The search direction is typically computed
according to
\begin{align}
  \label{eq:DO:gradients}
  s_k = \iteratesPE{H}{k}^{-1}\iteratesPE{g}{k}, \qquad \iteratesPE{g}{k} = \nabla_{\theta}
  \log p_{\theta}(y_{1:\T}) \big|_{\theta = \iteratesPE{\theta}{k}},
\end{align}
where $\iteratesPE{H}{k}$ denotes a positive definite matrix (e.g.\ the Hessian
$\nabla_{\theta}^2 \log p_{\theta}(y_{1:\T})$ or approximations thereof) adjusting the gradient
$\iteratesPE{g}{k}$.

\begin{figure}[p]%
  \includegraphics[width=\columnwidth]{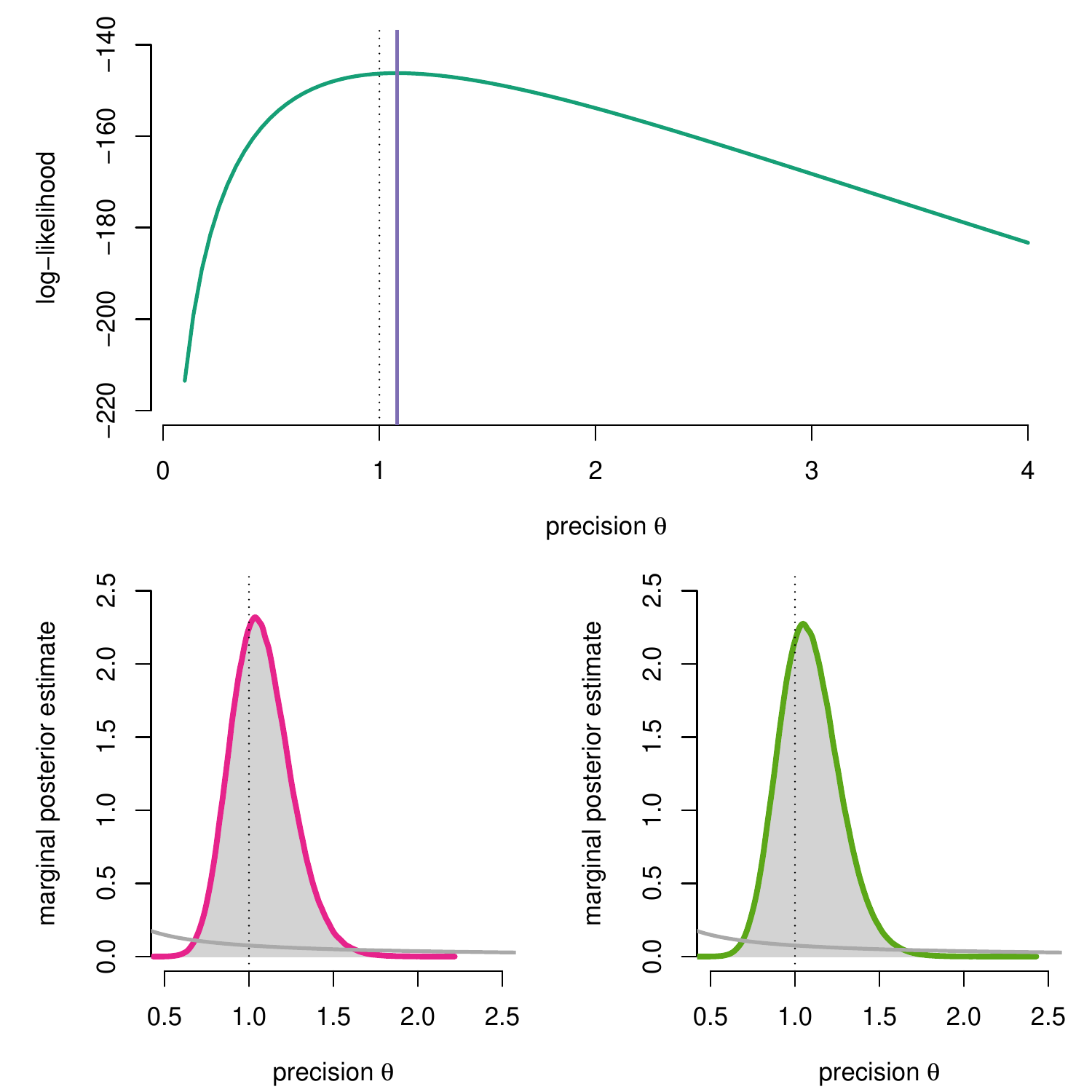}%
  \caption{\small Upper: the log-likelihood estimates (green) together with the ML parameter
    estimates of the LGSS model obtain by the DO and the EM algorithms, respectively. The estimates
    sit on top of each other and are shown in blue. Lower: parameter posterior estimates
    obtained by the MH (left) and the Gibbs (right) algorithms, respectively. The vertical dashed
    lines indicate the true parameters of the model from which the data is generated and the dark
    grey lines indicate the prior density.}%
  \label{fig:LGSS:ML}%
\end{figure}

\begin{example}[DO applied to the LGSS model] \label{example:LGSS:DO}%
  To apply the update in \eqref{eq:DO:ML} for estimating~$\theta$ in~\eqref{eq:LGSS}, we need to
  determine search direction~$s_k$ and the step lengths~$\alpha_k$. Here, we select the search
  direction as the gradient of the log-likelihood, \ie $H_k$ is the identity matrix. The
  log-likelihood for \eqref{eq:LGSS} can be expressed as
  \begin{align}
    \log p_{\theta}(y_{1:T}) = \sum_{t=2}^T
    \log \mathcal{N}(y_t; 0.5 \widehat{x}_{t|t-1}, P_{t|t-1} + 0.1),
    \label{eq:LGSSlikelihood}
  \end{align}
  where $\widehat{x}_{t|t-1}$ and $P_{t|t-1}$ denotes the predicted state estimate and its
  covariance obtained from a Kalman filter. The gradient $g_k$ in \eqref{eq:DO:gradients} of the
  log-likelihood \eqref{eq:LGSSlikelihood} can therefore be obtained by calculating
  $\nabla_{\theta} \widehat{x}_{t|t-1}$ and $\nabla_{\theta} P_{t|t-1}$, which can be obtained from
  the Kalman filter, using the so-called sensitivity derivatives introduced by
  \citet{Astrom:1980}. In the upper part of \fig{LGSS:ML}, we present the log-likelihood (blue)
  computed using the Kalman filter together with the estimate $\widehat{\theta}_{\text{ML}}$
  (orange) obtained by the DO algorithm.
\end{example}

Within the DO method the use of \smc arise in evaluation of the cost function in~\eqref{eq:DO:ML}
and its derivatives to compute the search directions~$\iteratesPE{s}{k}$ in~\eqref{eq:DO:gradients}.

\subsection{Bayesian identification via Metropolis--Hastings}
\label{sec:Marg:MH}
Let us now turn to the Bayesian formulation \eqref{eq:Bayes}. As above, to illustrate the idea we
consider first the simple case in which the marginalisation in~\eqref{eq:marg2} can be done exactly.
Still, in most nontrivial cases the posterior \pdf in~\eqref{eq:Bayes} cannot be computed in closed
form. The difficulty comes from the factor $p(y_{1:\T})$, known as the \emph{marginal likelihood},
which ensures that the posterior \pdf is properly normalised (\ie, that it integrates to one). The
marginal likelihood can be expressed as,
\begin{align}
  p(y_{1:\T}) 
  = \int p_\theta(y_{1:\T}) \pi(\theta)\myd\theta.
\end{align}
Even if the likelihood $p_\theta(y_{1:\T})$ is analytically tractable, we see that we need to carry
out an integration \wrt~$\theta$. 
Furthermore, computing a point estimate, say the posterior mean of $\theta$, also amounts to solving
an integral
\begin{align}
  \label{eq:2}
  \CExp{\theta}{y_{1:\T}} = \int \theta\, p(\theta \mid y_{1:\T})\myd \theta,
\end{align}
which may also be intractable.

A generic way of approximating such intractable integrals, in particular those
related to Bayesian posterior distributions, is to use a Monte Carlo method.  In
Section~\ref{sec:SMC} we will discuss, in detail, how \smc can be used to approximately
integrate over the \emph{latents states} of the system. 
The sequential nature of \smc makes it particularly well
suited for integrating over latent stochastic processes, such as the states in an \ssm.
However, to tackle the present problem of
integrating over the \emph{latent parameters} we shall consider a different class of methods denoted
as Markov chain Monte Carlo (\mcmc). 

The idea behind \mcmc is to simulate a Markov chain $\{\MCreal{\theta}{\iMC} \}_{\iMC\geq 1}$. The
chain is constructed in such a way that its \emph{stationary} distribution coincides with the
so-called \emph{target} distribution of interest, here $p(\theta\mid y_{1:\T})$.  If, in addition,
the chain is \emph{ergodic}---essentially meaning that spatial averages coincide with ``time''
averages---the sample path from the chain can be used to approximate expectations \wrt to the target
distribution:
\begin{align}
  \label{eq:EmpiricalApprPosterior}
  \frac{1}{M-k+1}\sum_{m=k}^{M} \varphi( \MCreal{\theta}{\iMC} ) \ConvAS \int \varphi(\theta) p(\theta \mid y_{1:\T}) \myd\theta,
\end{align}
as $M \goesto \infty$, for some test function $\varphi$. Here $\ConvAS$ denotes almost sure
convergence. Note that the first $k$ samples have been neglected in the estimator
\eqref{eq:EmpiricalApprPosterior}, to avoid the transient phase of the chain. This is commonly
referred to as the \emph{burn-in} of the Markov chain.

Clearly, for this strategy to be useful we need a systematic way of constructing the Markov chain with the desired properties.
One such way is to use the Metropolis--Hastings (\mh) algorithm.
The \mh algorithm uses an accept/reject strategy. At iteration $\iMC+1$, we propose a new sample $\MCpropState{\theta}$
according to,
\begin{align}
  \MCpropState{\theta} \sim \proposal(\cdot \mid \MCreal{\theta}{\iMC}),
  \label{eq:MH:proposal}
\end{align}
where $\proposal(\cdot)$ denotes a \emph{proposal} distribution designed by the user and
$\MCreal{\theta}{\iMC}$ denotes the sample from the previous iteration $\iMC$. The newly proposed
sample $\MCpropState{\theta}$ will then be added to the sequence (\ie $\MCreal{\theta}{\iMC+1} =
\MCpropState{\theta}$) with probability
\begin{subequations}
  \label{eq:MH:ap}
  \begin{align}
    \label{eq:MH:apa}
    \acceptProb &\triangleq 1 \wedge 
    \frac{p(\MCpropState{\theta}\mid y_{1:\T})}{ p(\MCreal{\theta}{\iMC} \mid y_{1:\T})}
    \frac{\proposal(\MCreal{\theta}{\iMC}\mid
      \MCpropState{\theta})}{\proposal(\MCpropState{\theta}\mid \MCreal{\theta}{\iMC})} \\
  \label{eq:MH:apb}
  &= 1 \wedge
  \frac{p_{\MCpropState{\theta}}(y_{1:\T})\prior(\MCpropState{\theta})}{p_{\MCreal{\theta}{\iMC}}(y_{1:\T})
    \prior(\MCreal{\theta}{\iMC})} \frac{\proposal(\MCreal{\theta}{\iMC}\mid
    \MCpropState{\theta})}{\proposal(\MCpropState{\theta}\mid \MCreal{\theta}{\iMC})},
  \end{align}
\end{subequations}
where $a\wedge b$ is used to denote $\min\left(a,b\right)$ and $\acceptProb$ is referred to as the
acceptance probability. Hence, with probability $1-\acceptProb$ the newly proposed sample is not
added to the sequence (\ie rejected) and in that case the previous sample is once more added to the
sequence $\MCreal{\theta}{\iMC+1} = \MCreal{\theta}{\iMC}$.


There exists a well established theory for the \mh algorithm, which for example establish that it is
ergodic and converges to the correct stationary distribution. This has been developed since the
algorithm was first introduced by \citet{Metropolis:1953} and
\citet{Hastings:1970}. \citet{NinnessH:2010} provides a nice account on the use of the \mh algorithm
for Bayesian system identification.


\begin{example}[MH applied to the LGSS model] \label{example:LGSS:MH}%
  To apply the MH algorithm for parameter inference in the LGSS model, we require a proposal
  distribution $q(\cdot)$ in~\eqref{eq:MH:proposal} and to calculate the acceptance
  probability~$\alpha$ in~\eqref{eq:MH:apb}. A standard choice for $q(\cdot)$ is a Gaussian random
  walk
  \begin{align}
    \label{eq:MH:LGSS:proposal}
    \proposal(\MCpropState{\theta} \mid \MCreal{\theta}{\iMC}) =
    \Npdf{\MCpropState{\theta}}{\MCreal{\theta}{\iMC}}{\sigma_q^2},
  \end{align}
  where $\sigma_q^2$ denotes the variance of the random walk. For this choice of proposal, the
  acceptance probability is
  \begin{align*}
    \acceptProb = 1 \wedge
    \frac{p_{\MCpropState{\theta}}(y_{1:\T})\prior(\MCpropState{\theta})}{p_{\MCreal{\theta}{\iMC}}(y_{1:\T})
    \prior(\MCreal{\theta}{\iMC})},
  \end{align*}
  where $q$ cancels since it is symmetric in~$\theta$. Note that the likelihood can be computed
  using the Kalman filter in analogue with~\eqref{eq:LGSSlikelihood} for the LGSS model. In the
  lower left part of \fig{LGSS:ML}, we present the resulting posterior estimate obtained from
  running the algorithm $M=20 \thinspace 000$ iterations (discarding the first $10 \thinspace 000$
  as burn-in).

\end{example}


\section{Identification strategy -- Data augmentation}
\label{sec:IdStrat:DA}
%
An alternative strategy to marginalisation is to make use of \emph{data augmentation}.  Intuitively,
we can think of this strategy as a systematic way of separating \emph{one} hard problem into
\emph{two} new and closely linked sub-problems, each of which is hopefully easier to solve than the
original problem. For the \ssm~\eqref{eq:SSM} these two problems amounts to
\begin{enumerate}
\item finding information about the state sequence $x_{1:\T}$.
\item finding information about the parameters $\theta$.
\end{enumerate}
The state sequence~$x_{1:\T}$ is thus treated as an \emph{auxiliary variable} that is estimated
together with the parameters~$\theta$. Using the data augmentation terminology
\citep{TannerW:1987,vanDykM:2001}, the state sequence $x_{1:\T}$ is referred to as the \emph{missing
  data}, as opposed to the \emph{observed data} $y_{1:\T}$. By augmenting the latter with the
former, we obtain the \emph{complete data} $\{x_{1:\T}, y_{1:\T}\}$.

Naturally, if the complete data were known, then identification of $\theta$ would have been much
simpler. In particular, we can directly write down the complete data likelihood
$p_{\theta}(x_{1:\T}, y_{1:\T})$ according to~\eqref{eq:jointXY}, which contrary to the observed
data likelihood $p_{\theta}(y_{1:\T})$, does not involve any marginalisation. The two likelihoods
are related via~\eqref{eq:marg}, suggesting that the complete data likelihood can indeed be used for
identifying the unknown parameters.

\subsection{Expectation Maximisation (\EM)}
\label{sec:EM}
%
Analogously with the marginalisation strategy, we can make use of data augmentation to address both
the frequentistic and the Bayesian identification problems.
For the former identification criteria, \eqref{eq:ML}, the result is the expectation maximisation
(\EM) algorithm \citep{DempsterLR:1977}. \EM provides an iterative procedure to compute \ml
estimates of unknown parameters $\theta$ in probabilistic models involving latent variables, like
the \ssm in~\eqref{eq:SSM}.

As a result of the conditional probability identity 
\begin{align*}
p_{\theta}(x_{1:\T}, y_{1:\T}) = p_{\theta}(x_{1:\T}\mid y_{1:\T}) p_{\theta}(y_{1:\T}),
\end{align*}
we can relate the observed and complete data log-likelihoods as
\begin{align}
  \label{eq:DA:EM:loglikelihoods}
  \log p_{\theta}(y_{1:\T}) = \log p_{\theta}(x_{1:\T}, y_{1:\T}) - \log p_{\theta}(x_{1:\T}\mid y_{1:\T}).
\end{align}
The \EM algorithms operates by iteratively maximising the \emph{intermediate quantity}
\begin{subequations}
  \label{eq:DA:EM:Qdef}
  \begin{align}
    \Q(\theta, \theta^{\prime}) &\triangleq \int \log p_{\theta}(x_{1:\T}, y_{1:\T})
    p_{\theta^{\prime}}(x_{1:\T} \mid y_{1:\T}) \myd x_{1:\T}\\
    &= \Expb{\theta^{\prime}}{\log p_{\theta}(x_{1:\T}, y_{1:\T}) \mid y_{1:\T}},
  \end{align}
\end{subequations}
according to:
\begin{itemize}
\item[(E)] $\Q(\theta, \theta[k]) = \Expb{\theta[k]}{\log p_{\theta}(x_{1:\T}, y_{1:\T}) \mid y_{1:\T}}$,
\item[(M)] $\theta[k+1] = \argmax{\theta\in\Theta} \Q(\theta, \theta[k])$.
\end{itemize}
We can show that iterating the above Expectation (E) and Maximisation (M) steps implies 
\begin{align}
  \label{eq:CA:EM:implication}
  \Q(\theta, \theta^{\prime}) \geq \Q(\theta^{\prime}, \theta^{\prime})
  \implies p_{\theta}(y_{1:\T}) \geq p_{\theta^{\prime}}(y_{1:\T}).
\end{align}
Hence, $\{\iterates{\theta}{k}\}_{k\geq 1}$ will by construction result in a monotonic
increase in likelihood values. Hence, the complete data log-likelihood
$\log p_{\theta}(x_{1:\T}, y_{1:\T})$ can, via the intermediate quantity $\Q$
in~\eqref{eq:DA:EM:Qdef}, be used as a surrogate for the original observed data likelihood function
$p_{\theta}(y_{1:\T})$ in solving the \ml problem~\eqref{eq:ML}. We are still required to compute
the integral~\eqref{eq:DA:EM:Qdef} and an important observation is now that we can approximate this
integral (and its derivatives \wrt $\theta$) 
using \smc.

\begin{example}[EM applied to the LGSS model] \label{example:LGSS:EM}%
  We need to compute the intermediate quantity to apply the EM algorithm for estimating the
  parameter in the LGSS model. For this model, we can write
  \begin{multline*}
    \Q(\theta, \theta^{\prime}) 
    = \text{const.} +
    \CExpb{\theta^{\prime}}{\log \mu_{\theta}(x_1)
    +
    \sum_{t=2}^T
    \log \mathcal{N}(x_{t};0.7x_{t-1},\theta^{-1})}{y_{1:\T}}
  \end{multline*}
  Note, that this expression results from that the parameter only is present in the latent state
  process.

  We can directly maximise the intermediate quantity for $\theta$ since the system is linear in the
  parameters. By taking the gradient of $\Q(\theta, \theta^{\prime})$, we obtain terms proportional
  to 
  $    E_{\theta^{\prime}} 
    \big[
    x_{t-1}
    ( x_t - 0.7 x_{t-1} )
    \mid
    y_{1:\T}
    \big]$, 
  where $\widehat{x}_{t|T}\widehat{x}_{t|T}$ and $\widehat{x}_{t|T}\widehat{x}_{t-1|T}$ denotes
  smoothed state estimates and $P_{t,t|T}$ and $P_{t-1,t|T}$ their covariances, respectively. These
  can be computed using a Kalman smoother and we refer the reader to \cite{GibsonN:2005} for the
  explicit expressions for implementing this. In the upper part of \fig{LGSS:ML}, we present the
  parameter estimate $\widehat{\theta}_{\text{ML}}$ (blue) obtained by the EM algorithm. We note
  that the parameter estimates obtained by the DO and EM algorithms are identical and
  overlapping. However, they differ from the true parameter due to the finite value of~$T$.
%
\end{example}
%

\subsection{Gibbs sampler}
The Gibbs sampler is an \mcmc method that produce samples from the joint distribution by
alternatively sampling from its conditionals. Let us consider the Bayesian formulation, with the
aim of computing~\eqref{eq:Bayes}. Inspired by the data augmentation strategy, start by assuming
that the complete data $\{x_{1:\T}, y_{1:\T}\}$ is available. Bayes' theorem then results in
\begin{align}
  \label{eq:DA:Gibbs:posterior}
  p(\theta\mid x_{1:\T}, y_{1:\T}) = \frac{p_{\theta}(x_{1:\T}, y_{1:\T})\prior(\theta)}{p(x_{1:\T}, y_{1:\T})}.
\end{align}
Intuitively, if the states~$x_{1:\T}$ were known, then the identification problem would be much
simpler and, indeed, computing the posterior in~\eqref{eq:DA:Gibbs:posterior} is typically 
easier than in~\eqref{eq:Bayes}.
Firstly, the complete data likelihood is provided in~\eqref{eq:jointXY}, whereas the likelihood
$p_\theta(y_{1:\T})$ is intractable in the general case. Secondly, in many cases of interest
it is possible to identify a prior for~$\theta$ that is conjugate to the \emph{complete data likelihood},
in which case the posterior~\eqref{eq:DA:Gibbs:posterior} is available in closed form. 



The problem with~\eqref{eq:DA:Gibbs:posterior} is of course that it hinges upon the state sequence being
known. However, assume that we can simulate a sample of the state trajectory~$x_{1:\T}$ from its conditional
distribution given the observed data $y_{1:\T}$ and the system parameters $\theta$, \ie from the \jsd.
Furthermore, assume that it is possible to sample from the distribution in \eqref{eq:DA:Gibbs:posterior}.
We can then implement the following algorithm: Initialise $\theta[0]\in\setTh$ arbitrarily and, for $\iMC \geq 0$,
\begin{subequations}
  \label{eq:DA:Gibbs}
  \begin{align}
    \label{eq:DA:Gibbsa}
    \text{Sample }&\MCreal{x_{1:\T}}{\iMC} \sim p_{ \MCreal{\theta}{\iMC} }(x_{1:\T}\mid y_{1:\T}). \\
    \label{eq:DA:Gibbsb}
    \text{Sample }&\MCreal{\theta}{\iMC+1} \sim p(\theta \mid \MCreal{x_{1:\T}}{\iMC}, y_{1:\T}).
  \end{align}
\end{subequations}
This results in the generation of the following sequence of random variables
\begin{align}
  \label{eq:MCGibbs}
  \MCreal{\theta}{0}, \MCreal{x_{1:\T}}{0},
  \MCreal{\theta}{1}, \MCreal{x_{1:\T}}{1},
  \MCreal{\theta}{2}, \MCreal{x_{1:\T}}{2},
  \dots,
\end{align}
which forms a mutual Markov chain in the parameters~$\theta$ and the states~$x_{1:\T}$,
$\{ \MCreal{\theta}{\iMC}, \MCreal{x_{1:\T}}{\iMC}\}_{\iMC\geq 1}$.
The procedure in~\eqref{eq:DA:Gibbs} represents a valid \mcmc method. More specifically, it is a particular
instance of the so-called \emph{Gibbs sampler}. The simulated Markov chain admits
the \emph{joint distribution} $p(\theta, x_{1:\T}\mid y_{1:\T})$ as a stationary distribution.
Furthermore, under weak conditions it can be can be shown to be ergodic.
That is, the Markov chain generated by the procedure \eqref{eq:DA:Gibbs} can be
used to estimate posterior expectations, and the estimators are consistent in the sense of
\eqref{eq:EmpiricalApprPosterior}.
%
%
Note that, if we are only interested in the
marginal distribution $p(\theta\mid y_{1:\T})$, then it is sufficient to store
the sub-sequence constituted by $\{\MCreal{\theta}{\iMC}\}_{\iMC\geq 1}$, obtained by simply discarding the samples
$\{\MCreal{x_{1:\T}}{\iMC}\}_{\iMC\geq 1}$ from~\eqref{eq:MCGibbs}. 

It is worth pointing out that it is possible to combine Gibbs sampling with other \mcmc methods. For instance,
if it is not possible to sample from posterior distribution \eqref{eq:DA:Gibbs:posterior} exactly,
then a valid approach is to replace step \eqref{eq:DA:Gibbsb} of the Gibbs sampler with, \eg,
an MH step with target distribution $p(\theta\mid x_{1:\T}, y_{1:\T})$. Similarly,
for nonlinear state space models, the \jsd in \eqref{eq:DA:Gibbsa} is not available in closed
form, but it is still possible to implement the strategy above by sampling the state trajectory
from an \mcmc kernel targeting the \jsd; see Section~\ref{sec:PGASBayesianSysId}.


\begin{example}[Gibbs applied to the LGSS model] \label{example:LGSS:Gibbs}%
  To implement the Gibbs sampler for parameter inference in the LGSS model, we need to sample from
  the conditional distributions in~\eqref{eq:DA:Gibbs}. To generate samples of state trajectories
  given $\theta$ and $y_{1:\T}$, we can make use of the factorisation%
  \begin{align}
    \label{eq:DA:Gibbs:bwdfactorization}
    p_\theta(x_{1:\T} \mid y_{1:\T}) = \left(\prod_{t=1}^{\T-1} p_\theta(x_t \mid x_{t+1},\ y_{1:t})\right) p_\theta(x_{\T} \mid y_{1:\T})
  \end{align}
  of the \jsd. Consequently, we can sample $\MCreal{x_{1:\T}}{\iMC}$ using the following
  \emph{backward simulation} strategy: Sample $\bx_\T \sim p_\theta(x_\T \mid y_{1:\T})$ and, for
  $t = \range{\T-1}{1}$, sample
  \begin{align}
    \label{eq:DA:Gibbs:LGSS:bwdKernel}
    \bx_{t} \sim p_\theta(x_t \mid \bx_{t+1},\ y_{1:t})
    \propto p_\theta(\bx_{t+1} \mid x_{t}) p_\theta(x_t \mid y_{1:t}).
  \end{align}
  We see that the backward simulation relies on the filtering distributions denoted by
  $\{p_\theta(x_t \mid y_{1:t}) \}_{t=1}^\T$ and, indeed, for the \lgss model we can obtain closed
  form expressions for all the involved densities by running a Kalman filter.

  In the second step, \eqref{eq:DA:Gibbsb}, we sample the parameters $\theta$ by
  \begin{subequations}
    \label{eq:DA:Gibbs:theta}
    \begin{align}
      \label{eq:DA:Gibbs:thetaa}
      \theta &\sim p(\theta \mid \widetilde x_{1:\T}, y_{1:\T}) = \gam(\theta \mid \alpha, \beta), \\
      \label{eq:DA:Gibbs:thetab}
      \alpha &= 0.01 + \frac{T}{2}, \\
      \label{eq:DA:Gibbs:thetac}
      \beta &= 0.01 + \frac{1}{2}\bigg(0.51 \widetilde x_1^2 + \sum_{t=1}^{T-1} (\widetilde x_{t+1} - 0.7 \widetilde x_t)^2 \bigg),
    \end{align}
  \end{subequations}
  which is the result of a standard prior-posterior update with a Gamma prior and Gaussian
  likelihood. The closed-form expression for $p(\theta \mid x_{1:\T}, y_{1:\T})$ is an effect of
  using a prior which is conjugate to the complete data likelihood.

  In the lower right part of \fig{LGSS:ML}, we present the resulting posterior
  estimate obtained from the Gibbs sampler with the same settings as for the MH
  algorithm. We note that the posterior estimates are almost identical for the two methods, which
  corresponds well to the established theory.

\end{example}

The data augmentation strategy (here implemented via the Gibbs sampler) enabled us to approximate
the posterior distribution $p(\theta\mid y_{1:\T})$ by separating the problem into two connected
sub-problems~\eqref{eq:DA:Gibbs}.


\section{Sequential Monte Carlo} 
\label{sec:SMC}
Sequential Monte Carlo (\smc) methods offer numerical
approximations to the state estimation problems associated with the nonlinear/non-Gaussian
\ssm~\eqref{eq:SSM}. The particle filter (\PF) approximates the filtering \pdf
$p_{\theta}(x_t\mid y_{1:t})$ and the particle smoother (\ps) approximates the joint smoothing \pdf
$p_{\theta}(x_{1:t}\mid y_{1:t})$, or some of its marginals. The \PF can intuitively be thought of
as the equivalent of the Kalman filter for nonlinear/non-Gaussian \ssm{s}.

The \smc approximation is an empirical distribution of the form
\begin{align}
  \label{eq:SMC:PFapprox}
  \widehat{p}_{\theta}(x_{t}\mid y_{1:t}) = \sum_{i = 1}^{\Np}\w_{t}^{i}\delta_{x_{t}^{i}}(x_{t}).
\end{align}
The samples $\{x_t^{i}\}_{i = 1}^{\Np}$ are often referred to as \emph{particles}---they are
point-masses ``spread out'' in the state space, each particle representing one hypothesis about the
state of the system. We can think of each particle $x_{t}^{i}$ as one possible system
state and the corresponding weight $\w_{t}^{i}$ contains information about how probable that
particular state is.

To make the connection to the marginalisation and data augmentation strategies introduced in the two
previous sections clear, we remind ourselves where the need for \smc arise in implementing 
these strategies to identify~$\theta$ in~\eqref{eq:SSM}.
The \PF is used to compute the cost function~\eqref{eq:DO:ML} and its derivatives in order to find
the search directions~\eqref{eq:DO:gradients}. To set up an \mh samler, we can make use of a
likelihood estimate provided by the \PF in order to compute the acceptance probabilities
in~\eqref{eq:MH:ap}. When it comes to data augmentation strategies, the \ps is used to approximate
the intermediate quantity in~\eqref{eq:DA:EM:Qdef} and in order to set up the Gibbs sampler,
particle methods are used to draw a realisation from the joint smoothing distribution
in~\eqref{eq:DA:Gibbsa}.

\subsection{Particle filter}
\label{sec:SMC:PF}
A principled solution to the nonlinear filtering problem is provided by the following two recursive
equations:
\begin{subequations}
  \label{eq:SMC:PF:tumu}
  \begin{align}
    \label{eq:SMC:PF:tumua}
    p_{\theta}(x_t\mid y_{1:t}) &= \frac{\g_{\theta}(y_t\mid x_t)p_{\theta}(x_t\mid y_{1:t-1})}{p_{\theta}(y_t\mid y_{1:t-1})}, \\
    \label{eq:SMC:PF:tumub}
    p_{\theta}(x_t \mid y_{1:t-1}) &= \int \f_{\theta}(x_t \mid x_{t-1})p_{\theta}(x_{t-1} \mid y_{1:t-1})\myd x_{t-1}.
  \end{align}
\end{subequations}
These equations can only be solved in closed form for very specific special cases, \eg, the \lgss
model which results in the Kalman filter. We will derive the particle filter as a general approximation
of~\eqref{eq:SMC:PF:tumu} for general nonlinear/non-Gaussian \ssm{s}. 

The particle filter---at least in its most basic form---can be interpreted as a sequential
application of \emph{importance sampling}. At each time step $t$ we use importance sampling to
approximate the filtering \pdf $p_{\theta}(x_t \mid y_{1:t})$. This is made possible by using the
expressions in \eqref{eq:SMC:PF:tumu} and by exploiting the already generated importance sampling
approximation of $p_{\theta}(x_{t-1} \mid y_{1:t-1})$.  At time $t=1$ we can find an empirical
distribution~\eqref{eq:SMC:PFapprox} by approximating 
$p_{\theta}(x_1 \mid y_1) \propto \g_{\theta}(y_1 \mid x_1) \mu_{\theta}(x_1)$ using importance
sampling in the normal sense.  We sample independently the particles $\{x_1^i\}_{i=1}^\Np$ from some
proposal distribution $\proposalSMC(x_1)$.  To account for the discrepancy between the proposal
distribution and the target distribution, the particles are assigned importance weights, given by
the ratio between the target and the proposal (up to proportionality), \ie
$w_1^i \propto \g_{\theta}(y_1 \mid x_1^i) \mu_{\theta}(x_1^i) / \proposalSMC(x_1^i)$, where the weights are
normalised to sum to one.

We proceed in an inductive fashion and assume that we have an empirical approximation of
the filtering distribution at time $t-1$ according to
\begin{align}
  \label{eq:SMC:PF:induction1}
  \widehat{p}_{\theta}(x_{t-1}\mid y_{1:t-1}) = \sum_{i = 1}^{\Np}\w_{t-1}^{i}\delta_{x_{t-1}^{i}}(x_{t-1}).
\end{align}
Inserting this into~\eqref{eq:SMC:PF:tumub} results in
\begin{align}
  \notag
  \widehat{p}_{\theta} (x_t \mid y_{1:t-1}) &= \int \f _{\theta} (x_t \mid x_{t-1}) \sum_{i = 1}^{\Np}
  \w_{t-1}^{i}\delta_{x_{t-1}^{i}}(x_{t-1}) \myd
  x_{t-1}\\
  \label{eq:SMC:PF:predictor_t}
  &= \sum_{i = 1}^{\Np}\w_{t-1}^{i}\f _{\theta} (x_t\mid x_{t-1}^{i}).
\end{align}
That is, we obtain a \emph{mixture distribution} approximating $p_{\theta}(x_t \mid y_{1:t-1})$,
where we have one mixture component for each of the $\Np$ particles at time $t-1$.  Furthermore,
inserting~\eqref{eq:SMC:PF:predictor_t} into~\eqref{eq:SMC:PF:tumua} results in the following
approximation 
of the filtering \pdf
\begin{align}
  \label{eq:SMC:PF:propto_filter_t}
  p_{\theta}(x_t\mid y_{1:t}) \approx \frac{ \g_{\theta}(y_t \mid x_t) }{ p_{\theta}(y_t\mid y_{1:t-1}) }
 \sum_{i = 1}^{\Np}\w_{t-1}^{i}  \f_{\theta}(x_t\mid x_{t-1}^{i}).
\end{align}
The task is now to approximate \eqref{eq:SMC:PF:propto_filter_t} using importance sampling.
 Inspired by the structure
of~\eqref{eq:SMC:PF:propto_filter_t} we choose a proposal density (again denoted by $\proposalSMC$)
of the same form, namely as a mixture distribution,
\begin{align}
  \label{eq:SMC:PF:mixture_proposal}
  \proposalSMC(x_t\mid y_{1:t}) \eqdef \sum_{i = 1}^{\Np}\apfw_{t-1}^{i} \proposalSMC (x_t \mid x_{t-1}^{i}, y_t),
\end{align}
where both the mixture components $\proposalSMC(x_t \mid x_{t-1}, y_t)$ and the mixture weights $\apfw_{t-1}$
are design choices constituting parts of the proposal distribution.

To generate a sample from the mixture distribution~\eqref{eq:SMC:PF:mixture_proposal} the following
two-step procedure is used; first we randomly select one of the components, and then we generate a
sample from that particular component. Note that we will sample $\Np$ particles from the proposal
distribution~\eqref{eq:SMC:PF:mixture_proposal}.
Let us use $a_t^{i}$ to denote the index of the mixture component selected for the $i^{\text{th}}$
particle at time $t$.  Now, since the probability of selecting component $\proposalSMC(x_t \mid
x_{t-1}^{j}, y_t)$ is encoded by its weight $\apfw_{t-1}^{j}$, we have that
\begin{align}
  \label{eq:SMC:PF:resampling}
  \myProb{a_t^{i} = j} &= \apfw_{t-1}^{j}, &j&=\range{1}{N}.
\end{align}
Subsequently, we can generate a sample from the selected component $a_t^{i}$ according to
$x_t^{i}\sim \proposalSMC(x_t\mid \resx_{t-1}^{i}, y_t)$, where $\resx_{t-1}^{i} \eqdef
x_{t-1}^{a_t^{i}}$. By construction $x_t^{i}$ is now a sample from the proposal
density~\eqref{eq:SMC:PF:mixture_proposal}.  The particle $\resx_{t-1}^{i}$ is referred to as the
ancestor particle of $x_t^{i}$, since $x_t^{i}$ is generated conditionally on
$\resx_{t-1}^{i}$. This also explains why the index~$a_t^{i}$ is commonly referred to as the
\emph{ancestor index}, since it indexes the ancestor of particle~$x_t^{i}$ at time~$t-1$.


In practice we sample the $\Np$ ancestor indices $\{ a_t^i \}_{i=1}^\Np$ according to
\eqref{eq:SMC:PF:resampling} in one go. This results in a new set of particles $\{ \resx_{t-1}^{i}
\}_{i=1}^{\Np}$ that are subsequently used to propagate the particles to time $t$. This procedure,
which (randomly) generates $\{ \resx_{t-1}^{i} \}_{i=1}^{\Np}$ by selection (sampling with
replacement) from among $\{ x_{t-1}^{i} \}_{i=1}^{\Np}$ according to some weights, is commonly
referred to as \emph{resampling}.

The next step is to assign importance weights to the new particles accounting for the discrepancy
between the target distribution $p_{\theta}(x_t\mid y_{1:t})$ and the proposal distribution
$\proposalSMC(x_t\mid y_{1:t})$. As before, the weights are computed as the ratio between the
(unnormalised) target \pdf and the proposal \pdf.  Direct use of~\eqref{eq:SMC:PF:propto_filter_t}
and~\eqref{eq:SMC:PF:mixture_proposal} results in
\begin{align}
  \label{eq:SMC:PF:weightFunction}
  \uw_t^i = \frac{ \g_{\theta}(y_t\mid x_t^i) \sum_{j = 1}^{\Np} \w_{t-1}^{j} \f_{\theta}(x_t^i \mid x_{t-1}^{j})  }
  { \sum_{j = 1}^{\Np} \apfw_{t-1}^{j} \proposalSMC(x_t^i \mid x_{t-1}^{j}, y_t) }.
\end{align}
By evaluating $\uw_t^i$ for $i = \range{1}{\Np}$ and normalising the weights, we
obtain a new set of weighted particles $\{x_t^i, \w_t^i \}_{i=1}^\Np$, constituting an empirical
approximation of $p_\theta(x_t\mid y_{1:t})$. This completes the algorithm, since these weighted particles
in turn can be used to approximate the filtering \pdf at time $t+1$, then at time $t+2$ and so on.

A problem with the algorithm presented above is that the weight calculation
in~\eqref{eq:SMC:PF:weightFunction} has a computational complexity of $\Ordo(\Np)$ for each
particle, rendering an overall computational complexity of $\Ordo(\Np^2)$, since the weights need to
be computed for all $\Np$ particles.  A pragmatic solution to this problem is to use the freedom
available in the proposal density and select it according to
$\proposalSMC(x_t\mid y_{1:t}) = \sum_{j = 1}^{\Np}\w_{t-1}^{j}\f_{\theta}(x_t\mid x_{t-1}^{j})$.
That is, we select ancestor particles with probabilities given by their importance weights and
sample new particles by simulating the system dynamics from time~$t-1$ to~$t$.
Inserting this into~\eqref{eq:SMC:PF:weightFunction} results in the simple expression
$\uw_t^i = \g_{\theta}(y_t\mid x_t^i)$, which brings the overall computational complexity down
to $\Ordo(\Np)$. The resulting algorithm is referred to as the \emph{bootstrap particle filter} and
it is summarised in Algorithm~\ref{alg:bPF}.

\begin{algorithm}[!b]
\caption{\textsf{Bootstrap particle filter (all operations are for $i=\range{1}{\Np}$)}}
\small
\begin{algorithmic}[1]
  \STATE \textbf{Initialisation ($t=1$): }
  \STATE \hspace{4mm} Sample $x_1^{i} \sim \mu_{\theta}(x_1)$.
  \STATE \hspace{4mm} Compute $\uw_1^{i} = \g_{\theta}(y_1\mid x_1^{i})$, normalise, $\w_1^{i} =
  \uw_1^{i}/\sum_{j=1}^{\Np}\uw_1^{j}$.
  \STATE \textbf{for} $t=2$ \textbf{to} $T$ \textbf{do}
  \STATE \hspace{4mm} \textbf{Resampling: } Sample $a_t^{i}$ with $\myProb{a_t^{i} = j} = \w_t^{j}$.
  \STATE \hspace{4mm} \textbf{Propagation: } Sample $x^{i}_{t} \sim \f_{\theta}(x_{t}\mid x_{t-1}^{a_t^{i}})$.
  \STATE \hspace{4mm} \textbf{Weighting: } Compute $\uw^{i}_{t} = \g_{\theta}(y_{t}\mid x_{t}^{i})$
    and normalise,$\vphantom{x_t^{a_t^i}}$\\
    \hspace{4mm} $\w^{i}_{t} = \uw^{i}_{t}/\sum_{j=1}^{\Np}\uw^{j}_{t}$.
  \STATE \textbf{end}
  \end{algorithmic}
  \label{alg:bPF}
\end{algorithm}

The bootstrap \PF was the first working particle filter, an early and influential derivation is
provided by \citet{Gordon:1993}. It is arguably the simplest possible implementation of \smc, but
nevertheless, it incorporates the essential methodological ideas that underpin the general \smc
framework. \emph{Importance sampling} (\ie propagation and weighting in Algorithm~\ref{alg:bPF}
above) and \emph{resampling} are used to sequentially approximate a sequence of probability
distributions of interest; here $\{p_{\theta}(x_t\mid y_{1:t})\}_{t\geq 1}$.

Selecting the dynamics as proposal distribution, as in the bootstrap particle filter, is appealing
due to the simplicity of the resulting algorithm. However, this choice is unfortunately also
suboptimal, since the current measurement $y_t$ is not taken into account when simulating the
particles $\{x_t^i \}_{i=1}^\Np$ from the proposal distribution.
A better strategy of reducing the computational complexity of the weight computation from
$\Ordo(\Np^2)$ to $\Ordo(\Np)$ is to \emph{target the joint distribution} of $(x_t, a_t)$ with an
importance sampler, instead of directly targeting the marginal distribution of $x_t$ as was done
above. Indeed, by explicitly introducing the ancestor indices as auxiliary variables in the importance sampler,
we obtain the weight expression
\begin{align}
  \uw_t^i = \frac{ \w_{t-1}^{a_t^i} \g_{\theta}(y_t\mid x_t^i)  \f_{\theta}(x_t^i \mid x_{t-1}^{a_t^i})  }
  {\apfw_{t-1}^{a_t^i}  \proposalSMC(x_t^i \mid x_{t-1}^{a_t^i}, y_t) },
\end{align}
as a more practical alternative to \eqref{eq:SMC:PF:weightFunction}. With this approach we have the
possibility of freely selecting the mixture weights $\apfw_{t-1}$ and mixture components $\proposalSMC(x_t \mid
x_{t-1}, y_t)$ of the proposal, while still enjoying an overall linear computational complexity.
The resulting algorithm is referred to as the auxiliary particle filter (\apf).  Rather than
providing the details of the derivation we simply refer to the original paper by \citet{PittS:1999} 
or our complete derivation in \citet{SchonL:2015}.

\subsection{Some useful properties of the \PF} \label{sec:SMC:propretiesPF}
As any Monte Carlo algorithm, the \PF can be interpreted as a \emph{random number generator}.
Indeed, the particles and the ancestor indices used within the algorithm are random variables,
and executing the algorithm corresponds to simulating a realisation of these variables.
It can be useful, both for understanding the properties of the \PF and for developing more advanced
algorithms around \smc, to make this more explicit.
%
%
Let
\begin{align}
  \label{eq:SMC:PF:allParticles}
  \mathbf{x}_t &\triangleq \crange{x_t^1}{x_t^\Np}, &&\text{and}& \mathbf{a}_t &\triangleq \crange{a_t^1}{a_t^\Np},
\end{align}
refer to all the particles and ancestor indices, respectively, generated by the \PF at time
$t$. The \PF in Algorithm~\ref{alg:bPF} then generates a \emph{single realisation} of a collection
of random variables $\{\mathbf{x}_{1:\T}, \mathbf{a}_{2:\T}\} \in \setX^{\Np\T} \times
\crange{1}{\Np}^{\Np(\T-1)}$.
Furthermore, since we know how these variables are generated, we can directly write down
their joint \pdf\footnote{\wrt to a natural product of Lebesgue and counting measure.} as,
\begin{align}
  \label{eq:SMC:extTarget}
  \psi_\theta(\mathbf{x}_{1:\T}, \mathbf{a}_{2:\T}\mid y_{1:\T}) \triangleq \prod_{i = 1}^\Np \mu_\theta(x_1^i) \prod_{t = 2}^{\T} \left \{ \prod_{i = 1}^\Np \w_t^{a_t^i} f_\theta(x_t^i \mid x_{t-1}^{a_t^i}) \right\}.
\end{align}
%
Naturally, any estimator derived from the \PF will also be a random variable.
From \eqref{eq:SMC:extTarget} we note that the distribution of this random variable
will depend on the number of particles $\Np$, and \emph{convergence} of the algorithm
can be identified with convergence, in some sense, of this random variable.
Specifically, let $\varphi: \setX \mapsto \R$ be some \emph{test function} of interest.
The posterior expectation $\CExpb{\theta}{\varphi(x_t)}{y_{1:t}} = \int \varphi(x_t) p_\theta(x_t \mid y_{1:t})\myd x_t$,
can be estimated by the PF by computing (\cf, \eqref{eq:EmpiricalApprPosterior}),
\begin{align}
  \label{eq:SMC:estimator}
  \widehat \varphi^N_t \eqdef \sum_{i=1}^N w_t^i \varphi(x_t^i).
\end{align}
There is a solid theoretical foundation for \smc, \eg, investigating the convergence of \eqref{eq:SMC:estimator}
to the true expectation as $\Np \goesto \infty$ and establishing non-asymptotic bounds on the approximation
error. 
%
The (types of) existing theoretical results are too numerous to be mentioned here,
and we refer to the book by \citet{DelMoral:2004} for a comprehensive treatment.
However, to give a flavour of the type of results that can be obtained we state a central limit
theorem (CLT) for the estimator \eqref{eq:SMC:estimator}.
Under weak regularity assumptions it holds that \citep{DelMoralM:2000,DelMoral:2004,Chopin:2004},
\begin{align}
  \label{eq:SMC:CLT}
  \sqrt{N} \left(\widehat \varphi^N_t - \CExpb{\theta}{\varphi(x_t)}{y_{1:t}} \right) \ConvDist \N(0, \sigma^2_t(\varphi)),
\end{align}
as $N\goesto\infty$ where $\ConvDist$ denotes convergence in distribution. The \emph{asymptotic estimator variance}
$\sigma^2_t(\varphi)$ depends on the test function $\varphi$, the specific PF implementation that is used
and, importantly, the properties of the state space model (an explicit expression for $\sigma^2_t(\varphi)$
is given by \citet{DoucetJ:2011}).

The CLT in \eqref{eq:SMC:CLT} is reassuring since it reveals that the estimator
converges 
at a rate $\sqrt{N}$, which is the same rate as for independent and identically distributed (\iid)
Monte Carlo estimators.  An interesting question to ask, however, is how the asymptotic variance
depends on $t$. In particular, recall from \eqref{eq:SMC:PF:propto_filter_t} that we use the
approximation of the filtering distribution at time $t-1$, in order to construct the \emph{target
  distribution},
which in turn is approximated by the particles at time $t$. This ``approximation of an
approximation'' interpretation of the PF may, rightfully, lead to doubts about the stability of the
approximation. In other words, will the asymptotic variance $\sigma^2_t(\varphi)$ grow exponentially
with~$t$?

Fortunately, in many realistic scenarios, the answer to this question is \emph{no}. The key to this
result is that the model exhibits some type of \emph{forgetting}, essentially meaning that the
dependence between states $x_s$ and $x_t$ diminishes (fast enough) as $|t-s|$ gets large. If this is
the case, we can bound $\sigma_t^2(\varphi) \leq C$ for some constant $C$ which is independent
of~$t$, ensuring the stability of the PF approximation.  We refer to
\citet{DelMoralG:2001,Chopin:2004} for more precise results in this direction.

In analogy with the Kalman filter, the PF does not only compute the filtering distribution,
but it also provides (an approximation of) the likelihood $p_\theta(y_{1:t})$,
which is central to the system identification problem.
For the bootstrap PF in
Algorithm~\ref{alg:bPF}, this is given by,
\begin{align}
  \label{eq:SMC:PF:likelihood}
  \widehat{p}_\theta(y_{1:t}) = \prod_{s=1}^{t}\left\{ \frac{1}{\Np}\sum_{i=1}^{\Np}\uw_s^{i} \right\}.
\end{align}
Note that the approximation is computed using the \emph{unnormalised} importance weights $\{\uw_s^i\}_{i=1}^N$.
The expression~\eqref{eq:SMC:PF:likelihood} can be understood by considering the factorisation~\eqref{eq:likelihood} and noting that the one-step predictive likelihood, by \eqref{eq:marg2},
can be approximated by,
\begin{align*}
  \widehat p_\theta(y_s \mid y_{1:s-1})
  &= \int g_\theta(y_s \mid x_s) \widehat p_\theta(x_s \mid y_{1:s-1})\myd x_s \\
  &= \frac{1}{N} \sum_{i=1}^N g_\theta(y_s \mid x_s^i)
  = \frac{1}{N} \sum_{i=1}^N \uw_s^i,
\end{align*}
where $\{x_s^i\}_{i=1}^N$ are simulated from the bootstrap proposal given by $r_\theta(x_s \mid y_{1:s}) = \widehat p_\theta(x_s \mid y_{1:s-1})$ (a similar likelihood estimator can be defined also for the general APF).

Sharp convergence results are available also for the likelihood estimator
\eqref{eq:SMC:PF:likelihood}.  First of all, the estimator is \emph{unbiased}, \ie
$\Expb{\psi_\theta}{ \widehat{p}_\theta(y_{1:t}) } = p_\theta(y_{1:t})$ for any value of $N$, where
the expectation is \wrt the randomness of the PF \citep{PittSGK:2012,DelMoral:2004}.  We will make
use of this result in the sequel. Furthermore, the estimator is convergent as $N\goesto\infty$. In
particular, under similar regularity and forgetting conditions as mentioned above, it is possible to
establish a CLT at rate $\sqrt{N}$ also for~\eqref{eq:SMC:PF:likelihood}. Furthermore, the
asymptotic variance for the normalised likelihood estimator can be bounded by $D\cdot t$ for some
constant $D$. 
Hence, in contrast with the filter estimator~\eqref{eq:SMC:estimator}, the asymptotic variance for
\eqref{eq:SMC:PF:likelihood} will grow with~$t$, albeit only linearly.
However, the growth can be controlled by selecting $N \propto t$, which provides a useful insight
into the tuning of the algorithm if it is to be used for likelihood estimation.

\subsection{Particle smoother}
\label{sec:SMC:PS}
The \PF was derived as a means of approximating the sequence of filtering densities
$\{p_{\theta}(x_t\mid y_{1:t})\}_{t\geq 1}$. We can also start from the forward smoothing relation
\begin{align}
  \label{eq:SMC:PS:fs}
  p_{\theta}(x_{1:t}\mid y_{1:t}) = p_{\theta}(x_{1:t-1}\mid y_{1:t-1})\frac{\f_{\theta}(x_t\mid x_{t-1})\g_{\theta}(y_t\mid x_t)}{p_{\theta}(y_t\mid y_{1:t-1})},
\end{align}
and derive the particle filter as a means of approximating the sequence of \emph{joint smoothing densities}
$\{p_{\theta}(x_{1:t}\mid y_{1:t})\}_{t\geq 1}$. Interestingly, the resulting algorithm is equivalent to
the \PF that
we have already seen. Indeed, by using the ancestor indices we can trace the
genealogy of the filter particles to get full state trajectories, resulting in the approximation
\begin{align}
  \label{eq:SMC:PSapprox}
  \widehat{p}_{\theta}(x_{1:t}\mid y_{1:t}) = \sum_{i = 1}^{\Np}\w_{t}^{i}\delta_{x_{1:t}^{i}}(x_{1:t}).
\end{align}
However, there is a serious limitation in using the \PF as a solution to the smoothing problem,
known as \emph{path degeneracy}. It arises due to the fact that the resampling step, by
construction, will remove particles with small weights and duplicate particles with high
weight. Hence, each resampling step will typically reduce the number of unique particles.  An
inevitable results of this is that for any given time $s$ there exists $t > s$ such that the \PF
approximation of $p_{\theta}(x_{1:t} \mid y_{1:t})$ collapses to a single particle at time~$s$.

One solution to the path degeneracy problem is to propagate information backwards in time, using a
forward/backward smoothing technique. The \jsd can be factorised as
in~\eqref{eq:DA:Gibbs:bwdfactorization} where each factor depends only on the \emph{filtering
  distribution} (\cf \eqref{eq:DA:Gibbs:LGSS:bwdKernel}).
Since the filter can be approximated without (directly) suffering from path degeneracy, this opens
up for a solution to the path degeneracy problem. An important step in this direction was provided
by \citet{GodsillDW:2004}, who made use of \emph{backward simulation} to simulate complete state
trajectories $\widetilde x_{1:\T}$, approximately distributed according to the joint smoothing
distribution $p_\theta(x_{1:\T} \mid y_{1:\T})$. The idea has since then been refined, see \eg
\citet{DoucGMO:2011,BunchG:2013}. Algorithms based on the combination of \mcmc and \smc introduced
by \citet{AndrieuDH:2010}, resulting in the particle \mcmc (\pmcmc) methods, also offer promising
solutions to the nonlinear state smoothing problem. For a self-contained introduction to particle
smoothers, see \citet{LindstenS:2013}.


\section{Marginalisation in the nonlinear \ssm} 
\label{sec:NL:marg}
Now that we have seen how \smc can be used to approximate the filtering distribution, as well as the
predictive and smoothing distributions and the likelihood, we are in the position of applying the
general identification strategies outlined in the previous sections to identify nonlinear/non-Gaussian state space
models.

\subsection{Direct optimisation using Fisher's identity}
\label{eq:Marg:FisherId}
Consider the maximum likelihood problem in~\eqref{eq:ML}.
The objective function, \ie the log-likelihood, can be approximated by \smc by
using \eqref{eq:SMC:PF:likelihood}.  However, many standard optimisation methods requires not only
evaluation of the cost function, but also the gradient and possibly the Hessian, in
solving~\eqref{eq:ML}. \smc can be used to compute the gradient via the use of Fisher's identity,
\begin{align}
  \nabla_{\theta} \log p_{\theta}(y_{1:\T}) \big|_{\theta = \iteratesPE{\theta}{k}} =
  \nabla_{\theta} \Q(\theta, \iteratesPE{\theta}{k}) \big|_{\theta = \iteratesPE{\theta}{k}},
  \label{eq:FishersIdentity}
\end{align}
where the intermediate quantity $\Q$ was defined in~\eqref{eq:DA:EM:Qdef}. It follows that
\begin{align}
  \nabla_{\theta} \log p_{\theta}(y_{1:\T}) =
  \Expb{\theta}{\nabla_{\theta}\log p_{\theta}(x_{1:\T}, y_{1:\T}) \mid y_{1:\T}}.
\end{align}
That is, the gradient of the log-likelihood can be computed by solving a smoothing problem.
This opens up for
gradient approximations via a particle smoother, as discussed in Section~\ref{sec:SMC:PS}; see \eg \citet{PoyiadjisDS:2011} for further
details. The Hessian can also be approximated using, for example, Louis' identity
\citep[\eg,][]{CappeMR:2005}.

Note that the gradient computed in this way will be stochastic, since it is approximated by an \smc method.
It is therefore common to choose a diminishing step-size sequence of the gradient ascent method according to
standard stochastic approximation rules; see \eg, \citet{KushnerY:1997,BenvenisteMP:1990}.
However, it should be noted that the approximation of the gradient of the log-likelihood will be \emph{biased} for a finite number of particles $\Np$, and
the identification method therefore relies on asymptotics in $\Np$ for convergence to a maximiser of~\eqref{eq:ML}.



\subsection{Using unbiased likelihoods within \mh}
\label{sec:PMH}
We can make use of the likelihood estimator~\eqref{eq:SMC:PF:likelihood} also for Bayesian
identification of nonlinear \ssm{s} via the \mh algorithm.
Indeed, an intuitive idea is to simply replace the intractable likelihood in the acceptance
probability~\eqref{eq:MH:ap} by the (unbiased) estimate $\widehat p_\theta(y_{1:\T})$.
What is maybe less intuitive is that this simple idea does in fact result in
a valid (in the sense that it has $p(\theta\mid y_{1:\T})$ as its stationary distribution) \mh
algorithm, for any number of particles $\Np\geq 1$. Let us now sketch why this is the case.


We start by introducing a (high-dimensional) auxiliary variable $u$ constituted by all the random
quantities generated by the \PF, \ie $u \eqdef \{ \bm{x}_{1:\T}, \bm{a}_{2:\T} \}$ distributed according
to $\psi_{\theta}(u\mid y_{1:\T})$ defined in~\eqref{eq:SMC:extTarget}. Note that the joint
distribution of the parameters $\theta$ and the auxiliary variables $u$,
\begin{subequations}
  \label{eq:PMH:IdealExtendedTarget}
  \begin{align}
    p(\theta, u\mid y_{1:\T}) &= \psi_{\theta}(u\mid y_{1:\T})p(\theta\mid y_{1:\T})\\
    \label{eq:PMH:IdealExtendedTargetb}
    &= \frac{ p_{\theta}(y_{1:\T}) \psi_{\theta}(u\mid y_{1:\T})\pi(\theta)}{p(y_{1:\T})},
  \end{align}
\end{subequations}
has the original target distribution $p(\theta\mid y_{1:\T})$ as one of its marginals. Inspired
by~\eqref{eq:PMH:IdealExtendedTargetb}, consider the following \emph{extended target} distribution
\begin{align}
  \label{eq:PMH:extTargetDist}
  \extTarget(\theta, u\mid y_{1:\T}) 
  = \frac{ \widehat{p}_{\theta, u}(y_{1:\T}) \psi_{\theta}(u\mid y_{1:\T})\pi(\theta)}{p(y_{1:\T})},
\end{align}
where we have made use of the unbiased likelihood estimate $\widehat{p}_{\theta, u}(y_{1:\T})$ from
the \PF (and indicate explicitly the dependence on $u$ in the notation for clarity).
We can now set up a \emph{standard \mh algorithm} that operates in the (huge)
\emph{non-standard extended space} $\setTh\times \setX^{\Np\T} \times \crange{1}{\Np}^{N(\T-1)}$
approximating the extended target distribution~\eqref{eq:PMH:extTargetDist}. The resulting algorithm
will generate samples asymptotically from $p(\theta\mid y_{1:\T})$ despite the fact that we employ an
\emph{approximate} likelihood in~\eqref{eq:PMH:extTargetDist}! To understand why this is the case,
let us marginalise~\eqref{eq:PMH:extTargetDist} \wrt the auxiliary variable~$u$:
\begin{align}
  \label{eq:PMH:marginaliseTarget}
  \int\extTarget(\theta, u\mid y_{1:\T})\myd u 
  = \frac{\pi(\theta)}{p(y_{1:\T})}\int \widehat{p}_{\theta, u}(y_{1:\T}) \psi_{\theta}(u\mid y_{1:\T})\myd u.
\end{align}
The fact that the likelihood estimate $\widehat{p}_{\theta,
  u}(y_{1:\T})$ produced by the \PF is unbiased means that
\begin{align}
  \label{eq:PMH:unbiasedLikelihood}
  \Expb{u|\theta}{\widehat{p}_{\theta, u}(y_{1:\T})} 
  = \int \widehat{p}_{\theta, u}(y_{1:\T}) \psi_{\theta}(u\mid y_{1:\T}) \myd u 
  = p_{\theta}(y_{1:\T}).
\end{align}
The marginalisation in~\eqref{eq:PMH:marginaliseTarget} can now be finalised, resulting in
$\int\extTarget(\theta, u\mid y_{1:\T})\myd u = p(\theta\mid y_{1:\T})$, proving that $p(\theta\mid y_{1:\T})$ is
recovered \emph{exactly} as the marginal of the extended target
distribution~\eqref{eq:PMH:extTargetDist}, despite the fact that we employed a \PF
\emph{approximation} of the likelihood using a finite number of particles $\Np$. This explains why
it is sometimes referred to as an \emph{exact approximation}. An interpretation is that using the
likelihood estimate from the \PF does change the marginal distribution \wrt~$u$
in~\eqref{eq:PMH:IdealExtendedTarget}, but it does \emph{not} change the marginal \wrt~$\theta$.

\begin{algorithm}[!t]
\caption{\textsf{Particle Metropolis Hastings (\pmh) for Bayesian system identification of nonlinear SSMs}}
\small
\begin{algorithmic}[1]
  \STATE Run a \PF (Algorithm~\ref{alg:bPF}) targeting $p(x_{1:\T} \mid \MCreal{\theta}{1})$ to
  obtain $\MCpropState{u} \sim \psi_{\MCreal{\theta}{1}}(u\mid y_{1:\T})$ and
  $\widehat{p}_{\MCreal{\theta}{1}, \MCpropState{u}}(y_{1:\T})$ according
  to~\eqref{eq:SMC:PF:likelihood}.
  \FOR{$\iMC=1$ to $M$}
  \STATE Sample $\MCpropState{\theta} \sim q(\cdot\mid \MCreal{\theta}{\iMC})$.

  \STATE Run a \PF (Algorithm~\ref{alg:bPF}) targeting $p(x_{1:\T} \mid \MCpropState{\theta})$ to
  obtain $\MCpropState{u} \sim \psi_{\MCpropState{\theta}}(u\mid y_{1:\T})$ and
  $\widehat{p}_{\MCpropState{\theta}, \MCpropState{u}}(y_{1:\T})$ according
  to~\eqref{eq:SMC:PF:likelihood}.

  \STATE Sample $d_{\iMC} \sim \mathcal{U}[0,1]$.
  \STATE Compute the acceptance probability $\alpha$ by~\eqref{eq:PMH:ap}.

  \IF{$d_{\iMC} < \alpha$}
  \STATE $\MCreal{\theta}{\iMC+1} \leftarrow \MCpropState{\theta}$ and
  $\widehat{p}_{\MCreal{\theta}{\iMC+1}}(y_{1:\T}) \leftarrow
  \widehat{p}_{\MCpropState{\theta}}(y_{1:\T})$.
  \ELSE
  \STATE $\MCreal{\theta}{\iMC+1} \leftarrow \MCreal{\theta}{\iMC}$ and $
  \widehat{p}_{\MCreal{\theta}{\iMC+1}}(y_{1:\T}) \leftarrow \widehat{p}_{
    \MCreal{\theta}{\iMC}}(y_{1:\T})$.
  \ENDIF
  \ENDFOR
\end{algorithmic}
\label{alg:PMH}
\end{algorithm}

Based on the current sample $(\MCreal{\theta}{\iMC}, \MCreal{u}{\iMC})$ a new sample
$(\MCpropState{\theta}, \MCpropState{u})$ is proposed according to
\begin{align}
  \label{eq:PMH:proposal}
  \MCpropState{\theta} \sim q(\cdot\mid \MCreal{\theta}{\iMC}), \qquad
  \MCpropState{u} \sim \psi_{\MCpropState{\theta}}(\cdot\mid y_{1:\T}).
\end{align}
We emphasise that simulation of $\MCpropState{u}$ corresponds to running a \PF with the model parameterised by $\MCpropState{\theta}$.
The probability of accepting the sample proposed in~\eqref{eq:PMH:proposal} as the next sample $(
\MCreal{\theta}{\iMC+1}, \MCreal{u}{\iMC+1})$ is given by
\begin{align}
    \acceptProb 
    \label{eq:PMH:ap}
    = 1 \wedge \frac{\widehat{p}_{\MCpropState{\theta}, \MCpropState{u}}(y_{1:\T}) 
      \pi(\MCpropState{\theta})}{\widehat{p}_{\MCreal{\theta}{\iMC},
        \MCreal{u}{\iMC}}(y_{1:\T}) \pi(\MCreal{\theta}{\iMC})} \, 
    \frac{q(\MCreal{\theta}{\iMC}\mid \MCpropState{\theta})}{q(\MCpropState{\theta}\mid \MCreal{\theta}{\iMC})},
\end{align}
which was obtained by inserting~\eqref{eq:PMH:extTargetDist} and~\eqref{eq:PMH:proposal}
into~\eqref{eq:MH:ap}.  In practice it is sufficient to keep track of the likelihood estimates
$\{ \widehat{p}_{\MCreal{\theta}{\iMC},\MCreal{u}{\iMC}} \}_{\iMC \geq 1}$, and we do not need to
store the complete auxiliary variable $\{\MCreal{u}{\iMC}\}_{\iMC \geq 1}$.  The above development
is summarised in Algorithm~\ref{alg:PMH}. It can be further improved by incorporating gradient and
Hessian information about the posterior into the proposal~\eqref{eq:PMH:proposal}, resulting in more
efficient use of the generated particles \citep{DahlinLS:2015}.

The \emph{particle Metropolis Hastings} algorithm constitutes one member of the \emph{particle \mcmc
  (\pmcmc)} family of algorithms introduced in the seminal paper by \citet{AndrieuDH:2010}. The
derivation above is along the lines of the pseudo-marginal approach due to
\citet{AndrieuR:2009}. The extended target construction~$\extTarget$, however, is the core of all \pmcmc
methods and they differ in that different (more or less standard) \mcmc samplers are used for this
(non-standard) target distribution. They also have in common that \smc is used as a proposal
mechanism on the space of state trajectories~$\setX^{\T}$.

\begin{figure}[p]%
  \includegraphics[width=\columnwidth]{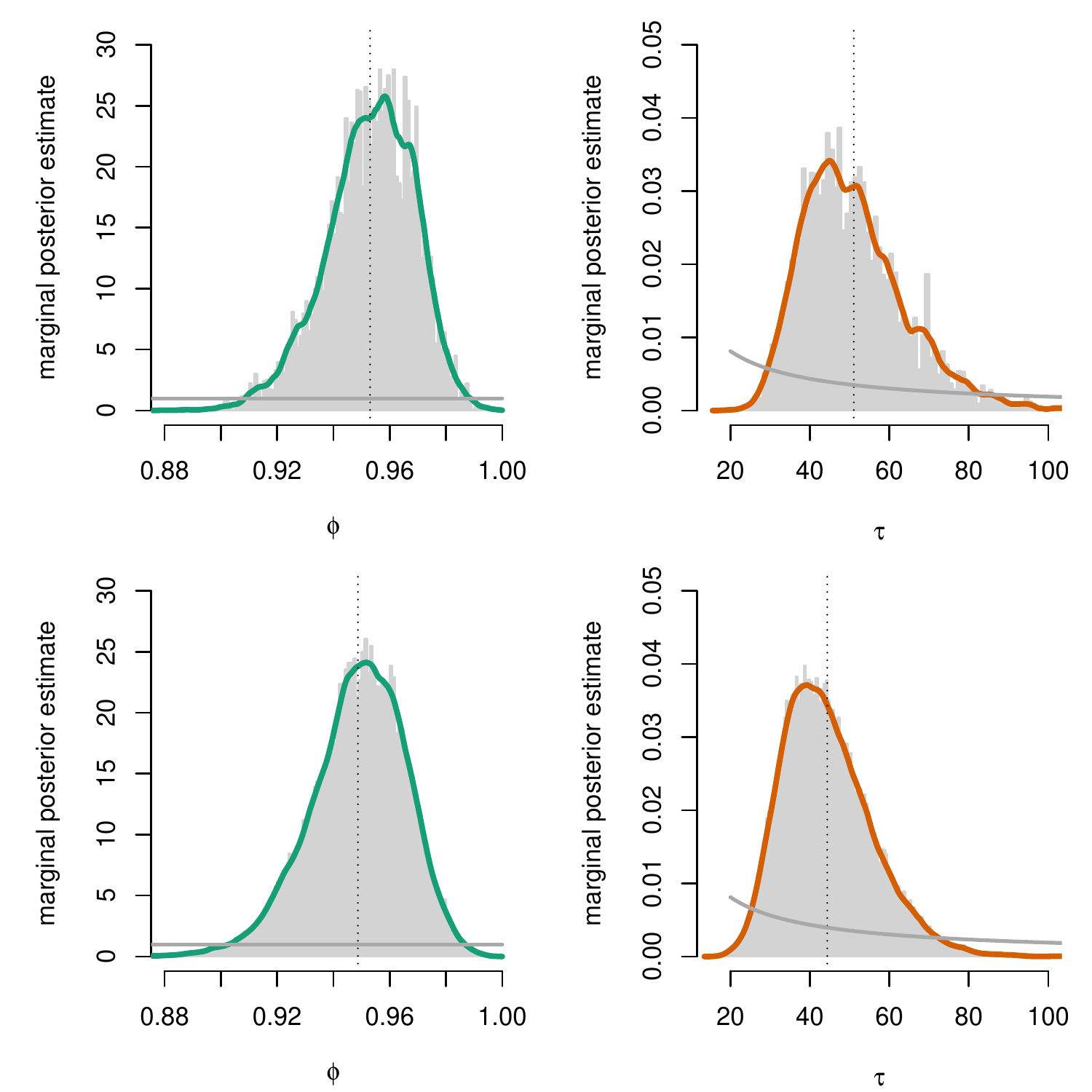}%
  \caption{\small The marginal posterior estimates for $\phi$ (left) and $\tau$ (right) using the PMH algorithm (upper) and the PGAS algorithm (lower). The dotted vertical and the dark grey lines indicate the estimated posterior mean and the prior densities, respectively.}%
  \label{fig:nlid-Bayesian}%
\end{figure}

\begin{example}[PMH applied to the NL-SSM]
  \label{example:Nonlinear:DA}%
  We make use of Algorithm~\ref{alg:PMH} to estimate the parameters in \eqref{eq:icevarvemodel}
  together with a simple Gaussian random walk,
  \begin{align*}
    \MCpropState{\theta}
    \sim 
    q(\cdot\mid \MCreal{\theta}{\iMC})
    = 
    \mathcal{N}( \MCreal{\theta}{\iMC}, 2.562^2 \Sigma /2 ),
  \end{align*}
  where $\Sigma$ denotes an estimate of the posterior covariance matrix. This choice is
  optimal for some target distributions as is discussed by \citet{SherlockThieryRobetsRosenthal2013}.
  The posterior covariance estimate is obtained as
  \begin{align*}
    \Sigma
    =
    10^{-5}
    \begin{bmatrix}
      22.51 & -4.53 \\ -4.53 & 2.57
    \end{bmatrix}
  \end{align*}
  using a pilot run of the algorithm. In the upper part of Figure~\ref{fig:nlid-Bayesian}, we present the resulting marginal posterior estimates. The posterior means $\widehat{\theta}_{\text{PMH}}=\{0.95,51.05 \}$ are
  indicated by dotted lines.
\end{example}

\section{Data augmentation in nonlinear \ssm}
\label{sec:NDA}
Algorithms implementing the data augmentation strategy treats the states as auxiliary variables that
are estimated along with the parameters, rather than integrating them out. Intuitively this results
in algorithms that alterate between updating $\theta$ and $x_{1:\T}$. 

\subsection{Expectation maximisation}
\label{sec:DAEM}
The expectation maximisation algorithm introduced in Section~\ref{sec:EM} separates the maximum
likelihood problem~\eqref{eq:ML} into two closely linked problems, namely the computation of the
intermediate quantity $\Q(\theta, \theta[k])$ and its maximisation \wrt $\theta$.
As previously discussed, computing the intermediate quantity corresponds to solving a smoothing problem.
Hence, for a nonlinear/non-Gaussian \ssm, a natural idea is to use a particle smoother, as discussed in Section~\ref{sec:SMC:PS},
to solve this subproblem.
The details of the algorithm are provided by 
\citet{CappeMR:2005,OlssonDCM:2008,SchonWN:2011}, whereas the general idea of making use of Monte Carlo integration to
approximate the E-step dates back to \cite{WeiT:1990}.

By this approach, a completely new set of simulated particles 
has to be generated at each iteration of the algorithm, since we continuously update the value of $\theta$.
Once an  approximation of $\Q(\theta, \theta[k])$ has been computed, the current 
particles are discarded and an entirely new set has to be generated at the next
iteration. While it does indeed result in a working algorithm it makes for an inefficient use of the
particles. The \pmcmc family of algorithms opens up for the construction of Markov kernels that can
be used to generate samples of the state trajectory (to be used in the
approximation of  $\Q(\theta, \theta[k])$)
in a computationally more efficient fashion, which serves as
one (of several) motivation of the subsequent development.

\subsection{Sampling state trajectories using Markov kernels}
\label{sec:DAsamplingMarkov}
%


We now introduce another member of the \pmcmc family of algorithms (recall \pmh from Section~\ref{sec:PMH})
that can be used whenever we are faced with the problem of sampling from an intractable \jsd
$p_{\theta}(x_{1:\T}\mid y_{1:\T})$. In those situations an exact sample can be replaced with a draw
from an MCMC kernel with stationary distribution $p_{\theta}(x_{1:\T}\mid y_{1:\T})$,
without introducing any systematic error, and \pmcmc opens up for using \smc to construct such MCMC kernels.


%
Here, we review a method denoted as particle Gibbs with ancestor sampling (PGAS), introduced by \citet{LindstenJS:2014}.
To construct the aforementioned Markov kernel, \pgas makes use of a procedure reminiscent of the \PF in Algorithm~\ref{alg:bPF}. The
only difference is that in \pgas we condition on the event that an \emph{a priori} specified 
state $x_{t}^{\prime}$ 
is always present in the particle system, for each time $t$.  Hence, the states
$\prange{x'_1}{x'_\T}$ must be retained throughout the sampling prodecure. To accomplish this we
sample $x_t^{i}$ according to the bootstrap \PF only for $i = \range{1}{\Np-1}$. The remaining
$\Np^{\text{th}}$ particle~$x_t^{\Np}$ is then set deterministically as $x_t^N = x_t^\prime$.  It is
often the case that we are interested in complete particles trajectories; \cf,
\eqref{eq:SMC:PSapprox}.  To generate a genealogical trajectory for $x_t^\prime$, it is possible to
connect it to one of the particles at time $t-1$, $\{x_{t-1}^i\}_{i=1}^\Np$ by sampling a value for
the corresponding ancestor index~$a_t^{\Np}$ from its conditional distribution.
This is referred to as \emph{ancestor sampling}, see Algorithm~\ref{alg:PGAS}. 

%
\begin{algorithm}[!b]
\caption{\textsf{\pgas kernel (with a bootstrap \PF)}}
\small
\begin{algorithmic}[1]
  \STATE \textbf{Initialisation ($t=1$): } Draw $x_1^{i} \sim \mu(x_1)$ for $i =
  \range{1}{\Np-1}$ and set $x_1^{\Np} = x_1^{\prime}$. Compute $\uw_1^i = \g_{\theta}(y_t\mid x_1^i)$ for $i =
  \range{1}{\Np}$.

  \FOR{$t=2$ to $\T$}
  \STATE Sample $a_t^{i}$ with $\myProb{a_t^{i} = j} = \w_{t-1}^{j}$ for $i = \range{1}{\Np-1}$.

  \STATE Sample $x^{i}_{t} \sim \f_{\theta}(x_{t}\mid x_{t-1}^{a_t^{i}})$ for $i = \range{1}{\Np-1}$.

  \STATE Set $x_t^{\Np} = x_t^{\prime}$.

  \STATE Draw $a_t^{\Np}$ with 
  $\myProb{a_t^{\Np} = j} \propto \uw_{t-1}^{j}
    \f_{\theta}(x_t^{\prime}\mid x_{t-1}^{j})$.
  \label{row:PGAS:as}


  \STATE Set $x_{1:t}^i = \{x_{1:t-1}^{a_t^i}, x_t^i\}$ for $i = \range{1}{\Np}$.

  \STATE Compute $\uw_t^i = \g_{\theta}(y_t\mid x_{t}^i)$ for $i = \range{1}{\Np}$.

  \ENDFOR
  \STATE Draw $k$ with $\myProb{k=i} \propto \uw_{\T}^{i}$.
  \STATE \textbf{Return} $x_{1:\T}^{\star} = x_{1:\T}^{k}$. 
\end{algorithmic}
\label{alg:PGAS}
\end{algorithm}

%
Note that Algorithm~\ref{alg:PGAS} takes as input a state trajectory $x_{1:\T}'  = \prange{x_1'}{x_\T'}$
and returns another state trajectory $x_{1:\T}^{\star}$, which is simulated randomly according to \emph{some} distribution (which, however, cannot be written on closed form).
Hence, we can view Algorithm~\ref{alg:PGAS} as sampling from a Markov kernel defined on the space of state trajectories $\setX^\T$.
This Markov kernel is referred to as the \pgas kernel.
The usefulness of the method comes from the fact that the \pgas kernel is a valid MCMC kernel for the \jsd $p_\theta(x_{1:\T} \mid y_{1:\T})$
for any number of particles $\Np \geq 2$!
A detailed derivation is provided by \citet{LindstenJS:2014}, who show that
the \pgas kernel is ergodic and that it admits the \jsd as its unique stationary distribution.
This implies that the state trajectories generated by \pgas can be used as samples from the \jsd. Hence, the
method is indeed an interesting alternative to other particle smoothers.
%
Moreover, the \pgas kernel can be used as a component in any (standard) \mcmc method.
In the subsequent section we will make explicit use of
this, both for \ml and Bayesian identification.


\section{Identification using Markov kernels}
\label{sec:IdMarkovKernel}

\subsection{Expectation maximisation revisited}
\label{sec:EMrevisited}
In Section~\ref{sec:DAEM} we made use of particle smoothers to approximate the intractable integral
defining the intermediate quantity $\Q(\theta, \theta[\iMC])$. 
However, it is possible to make more efficient use of the simulated variables by
using the \pgas Algorithm~\ref{alg:PGAS} and employing a \emph{stochastic
 approximation} update of the intermediate quantity $\Q$,
\begin{align}
  \label{eq:SAEM}
  \widehat{\Q}_{k}(\theta) = (1-\alpha_{k})\widehat{\Q}_{k-1}(\theta) + 
  \alpha_{k}\sum_{i=1}^{\Np} \w_{\T}^{i} \log p_{\theta}(x_{1:\T}^{i}, y_{1:\T}),
\end{align}
where $\alpha_{k}$ is the step size and $\{\w_{T}^{i}, x_{1:\T}^{i}\}_{i=1}^{\Np}$ is generated by Algorithm~\ref{alg:PGAS}. 
Stochastic approximation EM (SAEM) was introduced and analysed by \citet{DelyonLM:1999}
and it was later realised that it is possible to use MCMC kernels within SAEM \citep{AndrieuMP:2005}
(see also \citet{BenvenisteMP:1990}). The aforementioned \emph{particle SAEM} algorithm
for nonlinear system identification was presented by \citet{Lindsten:2013}
and it is summarised in Algorithm~\ref{alg:PGASmlSysId}.


\begin{algorithm}[!b]
\caption{\textsf{\pgas for ML sys. id. of nonlinear SSMs}}
\small
\begin{algorithmic}[1]
  \STATE \textbf{Initialisation: } Set $\theta[0]$ and $x_{1:\T}[0]$ arbitrarily. Set
  $\widehat{\Q}_{0} = 0$. 

  \FOR{$k\geq 1$}
  \STATE Run Algorithm~\ref{alg:PGAS} with $x_{1:\T}^\prime = x_{1:\T}[k-1]$. Set $x_{1:\T}[k] = x_{1:\T}^{\star}$.%

  \STATE Compute $\widehat{\Q}_{k}(\theta)$ according to~\eqref{eq:SAEM}.

  \STATE Compute $\theta[k] = \argmaxb{\widehat{\Q}_{k}(\theta)}$.

 \IF{convergence criterion is met} 
 \STATE{\textbf{return} $\theta[k]$} 
 \ENDIF

  \ENDFOR
\end{algorithmic}
\label{alg:PGASmlSysId}
\end{algorithm}

Note the important difference between the
\smc-based EM algorithm outlined in Section~\ref{sec:DAEM} and Algorithm~\ref{alg:PGASmlSysId}.
In the former we generate a completely new set of particles at each iteration,
whereas in particle \saem \emph{all} simulated particles contribute, but they are down-weighted using a
forgetting factor given by the step size.
This approach is more efficient in practice, since we can use much fewer particles at each iteration.
In fact, the method can be shown to converge to a maximiser of \eqref{eq:ML} even when using a fixed
number of particles $N\geq 2$ when executing Algorithm~\ref{alg:PGASmlSysId}.

  

\subsection{Bayesian identification}\label{sec:PGASBayesianSysId}
Gibbs sampling can be used to simulate from the posterior distribution~\eqref{eq:Bayes} or more
generally, the joint state and parameter posterior $p(\theta, x_{1:\T}\mid y_{1:\T})$. The \pgas
kernel allows us to sample the complete state trajectory $x_{1:\T}$ in one block. Due to the
invariance and ergodicity properties of the \pgas kernel, the validity of the Gibbs sampler is not
violated. We summarise the procedure in Algorithm~\ref{alg:PGASBayesianSysId}.

\begin{algorithm}[!t]
\caption{\textsf{\pgas for Bayesian sys. id. of nonlinear SSMs}}
\small
\begin{algorithmic}[1]
  \STATE \textbf{Initialisation: } Set $\theta[0]$ and $x_{1:\T}[0]$ arbitrarily. 

  \FOR{$\iMC=1$ to $M$}
  \STATE Run Algorithm~\ref{alg:PGAS} conditionally on $(x_{1:\T}{[\iMC-1]}, \theta[\iMC-1])$ and set $x_{1:\T}[\iMC] = x_{1:\T}^\star$.

  \STATE Draw $\theta[\iMC] \sim p(\theta\mid x_{1:\T}[\iMC], y_{1:\T})$.

  \ENDFOR
\end{algorithmic}
\label{alg:PGASBayesianSysId}
\end{algorithm}

\begin{example}[\pgas applied to~\eqref{eq:icevarvemodel}]
  \label{example:Nonlinear:DA}
  To make use of Algorithm~\ref{alg:PGASBayesianSysId} to estimate the parameters
  in~\eqref{eq:icevarvemodel}, we need to simulate from the conditional distribution
  $\theta[\iMC]~\sim~p(\theta\mid x_{1:\T}[\iMC], y_{1:\T})$. This distribution is not available in
  closed form, however we can generate samples from it by using rejection sampling with the
  following instrumental distribution
\begin{align*}
&q(\phi, \tau | x_{1:T}[\iMC], y_{1:T}) = \mathcal{G} \left( \tau; \alpha, \beta \right) \mathcal{N} \left( \phi; \tilde \mu, {\tilde \tau}^{-1} \right),\\
&\alpha = 0.01 + {T-1 \over 2}, \\
&\beta = 0.01 + {1 \over 2} \sum_{t=1}^T x_t[\iMC]^2 - {1 \over 2} \frac{\left( \sum_{t=1}^{T-1} x_{t+1}[\iMC] x_t[\iMC] \right)^2}{\sum_{t=2}^{T-1} x_t[\iMC]^2}, \\
&\tilde\mu = \frac{\sum_{t=1}^{T-1} x_{t+1}[\iMC] x_t[\iMC]}{\sum_{t=2}^{T-1} x_t[\iMC]^2}, ~\tilde\tau = \tau \sum_{t=2}^{T-1} x_t[\iMC]^2.
\end{align*}
In the lower part of Figure~\ref{fig:nlid-Bayesian}, we present the resulting marginal posterior estimates. The posterior means $\widehat{\theta}_{\text{PG}}=\{0.953,44.37 \}$ are indicated by dotted lines. 
\end{example}



\section{Future challenges}
We end this tutorial by pointing out directions for future research involving interesting challenges
where we believe that \smc can open up for significant developments.

Over two decades ago the \smc development started by providing a solution to the intractable
filtering problem inherent in the nonlinear \ssm. We have since then seen that \smc in indeed
much more widely applicable and we strongly believe that this development will continue, quite
possibly at a higher pace. This development opens up entirely new arenas where we can use \smc to
solve hard inference problems. To give a few concrete examples of this we have the Bayesian
nonparametric models (such as the Dirichlet and the Beta processes) that are extensively used in
machine learning. There are also the so-called \emph{spatio-temporal} models, which do not only have
structure in time, but also in a spatial dimension, imagine the weather forecasting problem. A known
deficiency of the standard (bootstrap) particle filter is its inability to handle high-dimensional
variables~$x_t$ \citep{bickelLB2008sharp}, which is usually the case in for example spatio-temporal
models. However, some very recent work has shown promising directions to tackle high-dimensional
models in a consistent way using \smc \citep{naessethls2014b,beskosCJKZ2014a,naessethLS2015nested}.

There is a well-known (but underutilised) duality between the control problem and the model learning
problem. Coupled with the various \smc based approximations this opens up for fundamentally new
controllers to be learnt by formulating the policy optimisation in control problems as an inference
problem. For some work along this direction, see \eg 
\citep{DoucetJT:2010,HoffmanKdFD:2009,ToussaintS:2006}. 

The \pmcmc family of methods that have been discussed and used throughout this tutorial is a
concrete example of another interesting trend, namely that of coupling various sampling methods into
more powerful solutions. This is a trend that will continue to evolve. The online (Bayesian)
inference problem is also a future challenge where we believe that \smc will play an important role.

\appendix
\section{Implementation details}
This appendix provides additional implementation details and clarifications about the numerical illustrations
given in the paper.

\subsection{Linear Gaussian state space model}
The \lgss model studied in Example~\ref{example:LGSS} is given by:
\begin{subequations}
  \label{eq:app:LGSS}
  \begin{align}
    \label{eq:app:LGSSa}
    x_{t+1} &= 0.7x_t + v_t, \quad &v_t& \sim \N(0,\theta^{-1}),\\
    \label{eq:app:LGSSb}
    y_t &= x_t + e_t, \quad &e_t& \sim \N(0,0.1),\\
    \label{eq:app:LGSSc}
    (\theta &\sim \gam(0.01, 0.01) ),
  \end{align}
\end{subequations}
where the unknown parameter $\theta$ corresponds to the \emph{precision} of the process noise $v_t$
(\ie, $\theta^{-1}$ is the process noise variance).  Note that the prior for the Bayesian model is
chosen as the Gamma ($\gam$) distribution with know parameters for reasons of simplicity (it
provides positive realizations and it is the conjugate prior). Specifically, $\gam(a,b)$ denotes a
Gamma distribution with shape $a$ and rate $b$ such that the mean is $a/b$:
\begin{align}
  \label{eq:app:lgss:prior}
  \gam(\theta ; a,b) = \frac{ b^a \theta^{a-1} \exp(-b\theta) }{ \Gamma(a) }.
\end{align}
The state process is assumed to be stationary. This implies that the distribution of the initial state
(\ie the state at time $t=1$) is given by,
\begin{align*}
  p(x_1 \mid \theta) = \N(x_1 ; 0, \{ (1-0.7^2)\theta \}^{-1} ) = \N(0, \{ 0.51 \theta \}^{-1} ).
\end{align*}
Identification of $\theta$ is based on a simulated data set consisting of $\T = 100$ samples
$y_{1:100}$ with true parameter $\theta_0=1$.

\subsubsection{Marginalization}

\paragraph{Direct optimization}

The log-likelihood for the \lgss model is given by
\begin{align}
  \nonumber
V(\theta) &= \log p_{\theta}(y_{1:T}) = \log \prod_{t=1}^T p_{\theta}(y_t \mid y_{1:t-1}) = \sum_{t=1}^T \log p_{\theta}(y_t \mid y_{1:t-1}) \\
  \nonumber
  &= \sum_{t=1}^T \log \N(y_t ;  \widehat{x}_{t \mid t-1}, \underbrace{ P_{t \mid t-1}  + 0.1}_{\triangleq \Lambda_t}) \\
  \nonumber
  &= \sum_{t=1}^T \bigg[-\frac{1}{2} \log 2 \pi -\frac{1}{2} \log \Lambda_t - \frac{1}{2\Lambda_t} (y_t -  \widehat{x}_{t \mid t-1})^2 \bigg] \\
  \label{eq:app:lgss:loglik}
  &= -\frac{T}{2} \log 2 \pi -\frac{1}{2} \sum_{t=1}^T \bigg[ \log \Lambda_t + \frac{1}{\Lambda_t} (y_t -  \widehat{x}_{t \mid t-1})^2 \bigg] 
\end{align}
where $\widehat{x}_{t\mid t-1}$ is the optimal predictor and $P_{t \mid t-1}$ is the covariance of the prediction error.
These quantities can be computed by the Kalman filter via the following recursions:
\begin{subequations} \label{eq:app:kalmanfilter}
\begin{align}
  \Lambda_t &=  P_{t \mid t-1}  + 0.1 \\
  K_t &= 0.7 P_{t \mid t-1}  \Lambda_t^{-1} \label{eq:app:kalmangain}\\
  \widehat{x}_{t+1 \mid t} &= 0.7 \widehat{x}_{t \mid t - 1} + K_t(y_t -  \widehat{x}_{t \mid t - 1}) \label{eq:app:kalmanpred}\\
  P_{t+1 \mid t} &= 0.49 P_{t \mid t-1} + \theta^{-1} - 0.7 K_t  P_{t \mid t-1} \label{eq:app:kalmanpredcov}
\end{align}
\end{subequations}
initialized with $\widehat{x}_{1 \mid 0} = 0$ and $P_{1 \mid 0} = (0.51\theta)^{-1}$, the mean and covariance
of $x_1$, the state at time $t=1$.

The gradient of the objective function becomes
\begin{align*}
  \frac{d}{d \theta} V(\theta)
    &= -\frac{1}{2} \sum_{t=1}^T \bigg[\frac{d}{d\theta} \log \Lambda_t + \frac{d}{d\theta} \frac{1}{\Lambda_t} (y_t -  \widehat{x}_{t \mid t-1})^2\bigg] \\
    &= -\frac{1}{2}\sum_{t=1}^T \bigg[ \frac{1}{\Lambda_t} \frac{d\Lambda_t}{d\theta} -  \frac{2}{\Lambda_t} (y_t - \widehat{x}_{t \mid t-1}) \frac{d \widehat{x}_{t \mid t-1}}{d \theta} - \frac{1}{\Lambda_t^2} (y_t -  \widehat{x}_{t \mid t-1})^2 \frac{d \Lambda_t}{d\theta}  \bigg],
\end{align*}
where
\begin{align*}
\frac{d \Lambda_t}{d\theta} &=  \frac{d P_{t \mid t-1}}{d\theta}.
\end{align*}
In order to compute the gradient, we need to compute $\frac{d}{d\theta} \widehat{x}_{t \mid t-1}$ and $\frac{d}{d\theta} P_{t \mid t-1}$.
This can be done recursively by differentiating \eqref{eq:app:kalmanfilter} with respect to $\theta$.
We get
\begin{subequations}
\begin{align}
\frac{\myd K_t}{\myd\theta} 
    &=\frac{0.7}{\Lambda_t}  \left( 1- \frac{ P_{t \mid t-1}}{\Lambda_t} \right) \frac{\myd P_{t \mid t-1}}{\myd\theta}, \\ 
 \frac{\myd \widehat{x}_{t+1 \mid t}}{\myd \theta} &= (0.7 - K_t) \frac{\myd \widehat{x}_{t \mid t-1}}{\myd\theta} + (y_t -  \widehat{x}_{t \mid t-1}) \frac{\myd K_t}{\myd\theta}, \\
  \frac{\myd P_{t + 1 \mid t}}{\myd\theta} &= (0.49 - 0.7K_t ) \frac{\myd P_{t \mid t-1}}{\myd\theta} - \frac{1}{\theta^2} - 0.7 P_{t \mid t-1} \frac{\myd K_t}{\myd\theta} .
\end{align}
\end{subequations}
For a more complete treatment of this problem, see \citep{Astrom:1980}.


\paragraph{Metropolis Hastings}
The first task in setting up an \mh algorithm to compute $p(\theta\mid y_{1:\T})$ is to decide on
  which proposal distribution to use. For simplicity, let us make use of a random walk
\begin{align}
  \label{eq:app:MH:LGSS:proposal}
  \proposal(\MCpropState{\theta} \mid \MCreal{\theta}{\iMC}) =
  \Npdf{\MCpropState{\theta}}{\MCreal{\theta}{\iMC}}{0.1}.
\end{align}
Now that we can propose new samples according to~\eqref{eq:app:MH:LGSS:proposal}, the
probability of accepting the new sample $\MCpropState{\theta}$ has to be
computed. The prior $\prior(\cdot)$ and the proposal $\proposal(\cdot)$ are given
by~\eqref{eq:app:LGSSc} and~\eqref{eq:app:MH:LGSS:proposal}, respectively. Hence, it remains to compute the
likelihood. The log-likelihood, denoted by $V(\theta)$ (to agree with the previous section),
is given by \eqref{eq:app:lgss:loglik} and it can be computed by running the Kalman filter \eqref{eq:app:kalmanfilter}.
The resulting expression for the acceptance probability is thus given by
\begin{align}
  \nonumber
  \acceptProb &= 1 \wedge 
    \frac{p_{\MCpropState{\theta}}(y_{1:T}) \prior(\MCpropState{\theta}) \proposal(\MCreal{\theta}{\iMC}\mid \MCpropState{\theta})}
    {p_{\MCreal{\theta}{\iMC}}(y_{1:T}) \prior(\MCreal{\theta}{\iMC}) \proposal(\MCpropState{\theta}\mid \MCreal{\theta}{\iMC})} \\
    \nonumber
    &= 1 \wedge  \frac{p_{\MCpropState{\theta}}(y_{1:T}) \prior(\MCpropState{\theta}) }
    {p_{\MCreal{\theta}{\iMC}}(y_{1:T}) \prior(\MCreal{\theta}{\iMC}) } \\
  \label{eq:app:MH:LGSS:ap}
    &= 1 \wedge \exp( V(\MCpropState{\theta}) - V( \MCreal{\theta}{\iMC} )
    - 0.99 \log(\MCpropState{\theta} / \MCreal{\theta}{\iMC}) - 0.01( \MCpropState{\theta} - \MCreal{\theta}{\iMC})), 
\end{align}
where the first equality follows from the fact that the random walk proposal \eqref{eq:app:MH:LGSS:proposal} is
symmetric in $ \MCpropState{\theta}$ and $\MCreal{\theta}{\iMC}$, and the second equality follows from \eqref{eq:app:lgss:prior} and~\eqref{eq:app:lgss:loglik}.
Note also that the prior $\pi$ is only supported on positive values on $\theta$, so if a negative value is proposed it is automatically rejected ($\alpha = 0$).




\subsubsection{Data augmentation}

\paragraph{Expectation Maxmimisation}
In the problem setting, $x_{1:T}$ act as the latent variables. In order to apply the EM algorithm, we need to calculate the following surrogate function
\begin{align}
\label{eq.q}
\Q(\theta, \iterates{\theta}{k}) = 
 \mathbb{E}_{\iterates{\theta}{k}}\log p_\theta(x_{1:T},y_{1:T}).
\end{align}
Expanding the right hand side of Eq. \eqref{eq.q} gives that
\begin{align*}
&\Q(\theta, \iterates{\theta}{k}) = \mathbb{E}_{\iterates{\theta}{k}}\!\left( \log p_\theta(x_1)p(y_1|x_1)\prod_{t=2}^{t=T} p_\theta(x_{t}|x_{t-1})p(y_t|x_t) \right) \\
& = \mathbb{E}_{\iterates{\theta}{k}}\! \left( \log p_\theta(x_1) + \sum_{t=2}^{T}\log p_\theta(x_{t}|x_{t-1}) + \sum_{t=1}^{T} \log p(y_t|x_{t}) \right)
\end{align*}
In the following, we will drop the terms which are independent of $\theta$ 
and use the linearity of the expectation operator,
which gives that
\begin{align*}
\Q(\theta, \iterates{\theta}{k})
 = \frac{1}{2} \left\{ T \log(\theta) - \theta \left[0.51 \mathbb{E}_{\iterates{\theta}{k}} (x_1^2) + \sum_{t=1}^{T-1}\mathbb{E}_{\iterates{\theta}{k}}\! \left((x_{t+1}-0.7x_{t})^2 \right)\right] \right\} + \text{const.}
\end{align*}
We see that we need to compute expectations \wrt the smoothing distribution (for the model parameterised by $\iterates{\theta}{k}$),
which can be done by running any convenient Kalman smoother.


Next, the M step amounts to maximising $\Q(\theta, \iterates{\theta}{k})$ with respect to $\theta$. In our case, this
maximisation has a closed form solution, given by
\begin{align}
\iterates{\theta}{k+1} = \frac{T}{0.51 \mathbb{E}_{\iterates{\theta}{k}} (x_1^2) + \sum_{t=1}^{T-1}\mathbb{E}_{\iterates{\theta}{k}}\! \left((x_{t+1}-0.7x_{t})^2 \right)}.
\end{align}

\paragraph{Gibbs} The Gibbs sampler iterates between simulating $x_{1:T}$ from $p_\theta(x_{1:T} \mid  y_{1:T})$ and
$\theta$ from $p(\theta \mid x_{1:T}, y_{1:T})$. Simulating $x_{1:T}$ is done by \emph{backward sampling} as shown in Eq.\ \eqref{eq:DA:Gibbs:LGSS:bwdKernel}.
The filtering densities $p_\theta(x_t \mid y_{1:t}) = N(x_t \mid \widehat x_{t|t}, P_{t|t})$ are computed by running a Kalman filter. We then obtain the following
expression for the backward kernel:
\begin{align*}
  p_\theta(x_t \mid y_{1:t}, \widetilde x_{t+1}) = N(x_t \mid \mu_t, \Sigma_t),
\end{align*}
with
\begin{align*}
  \mu_t &= \widehat x_{t|t} + 0.7 P_{t|t} \left( \frac{1}{\theta} + 0.49 P_{t|t} \right)^{-1} (\widetilde x_{t+1} - 0.7 \widehat x_{t|t}), \\
  \Sigma_t &= P_{t|t} - 0.49 P_{t|t}^2 \left( \frac{1}{\theta} + 0.49 P_{t|t} \right)^{-1}.
\end{align*}
As for the conditional distribution of $\theta$: due to the fact that the
Gamma distribution is the conjugate prior for the precision in a Gaussian model, we obtain
a closed form expression for,
\begin{align*}
&p(\theta \mid x_{1:T}, y_{1:T}) = p(\theta \mid x_{1:T}) \propto p(x_{1:T} \mid \theta) p(\theta) = p(\theta) p(x_1 \mid \theta) \prod_{t=1}^{T-1}p(x_{t+1} \mid x_t, \theta) \\
  &\quad\propto \theta^{a-1} \exp({-b \theta}) \sqrt{\theta} \exp\bigg(-\frac{0.51\theta}{2}x_1^2\bigg)  \times\prod_{t=1}^{T-1} \sqrt{\theta} \exp \bigg( -\frac{\theta}{2}(x_{t+1} - 0.7x_t)^2 \bigg) \\
  &\quad= \theta^{a+\frac{T}{2}-1} \exp\Bigg({-\bigg(b} +\frac{0.51}{2}x_1^2 + \frac{1}{2} \sum_{t=1}^{T-1} (x_{t+1} - 0.7x_t)^2 \bigg) \theta \Bigg)  \\
  &\quad \propto \gam\Bigg(\theta ; a + \frac{T}{2}, b + \frac{1}{2}\bigg(0.51x_1^2 + \sum_{t=1}^{T-1}(x_{t+1} - 0.7 x_t)^2 \bigg)\Bigg).
\end{align*}
Note that we have use proportionality, rather than equality, in several of the steps
above. However, since we know that $p(\theta \mid x_{1:T}, y_{1:T})$ is a PDF (\ie, it integrates to one), it is sufficient to obtain an expression which is proportional
to a Gamma PDF (the penultimate line). By normalisation we then obtain that 
$p(\theta \mid x_{1:T}, y_{1:T})$ is indeed given by the Gamma PDF on the final line.

The above derivation can straightforwardly be generalized to derive a Gibbs sampler for a general 
\lgss model, the details are provided by \citet{WillsSLN:2012}.

\subsection{Nonlinear example}
Example~\ref{example:Nonlinear} is borrowed from \cite{ShumwayS:2011} (see pages 63, 131, 151, 270, 280). Consider a
data set consisting of $634$ measurements of the thickness of ice varves (the layers of clay
collected in glaciers) formed at a location in Massachusetts between years $9,883$ and $9,250$
BC. The data is modelled using a nonlinear state space model given by,
\begin{subequations}
  \begin{align}
    x_{t+1}|x_t &\sim \N(x_{t+1}; \phi x_t, \tau^{-1} ), \\
    y_{t}|x_t   &\sim \gam(y_t; 6.25, 0.256 \exp(-x_t) ),
  \end{align}%
  \label{eq:app:icevarvemodel}%
\end{subequations}%
\noindent with the parameters $\theta=(\phi, \tau)^\+$.
The system is assumed to be stable and the state process stationary. This implies that the distribution of the initial state
(\ie the state at time $t=1$) is given by,
\begin{align*}
  p(x_1 \mid \theta) = \N(x_1 ; 0, \{ (1-\phi^2)\tau \}^{-1} ).
\end{align*}
In the Bayesian setting, we use a uniform prior for $\phi$ to reflect the stability
assumption, and a conjugate Gamma prior for $\tau$:
\begin{align*}
  p(\phi) &= \uni(\phi ; -1, 1), \\
  p(\tau) &= \gam(\tau ; 0.01, 0.01).
\end{align*}


Identification of $\theta$ is based on the measured data set consisting of $\T = 634$ samples $y_{1:634}$.


\subsubsection{Marginalization}


\begin{algorithm}[!t]
\caption{\textsf{Gradient-based maximum likelihood inference in NL-SSMs}}
\begin{footnotesize}
\textsc{Inputs:} $K$ (no.\ iterations), $y_{1:T}$ (data), $\theta_0$ (initial parameter), $\gamma$ and $\alpha$ (step length sequence). \\
\textsc{Outputs:} $\widehat{\theta}$ (est.\ of parameter).
\algrule[.4pt]
\begin{algorithmic}[1]
	\STATE Initialise the parameter estimate $\widehat{\theta}_0 = \theta_0$ and set $k=1$.
	\WHILE{$k \leq N$ or until convergence}
		\STATE Run the FFBSi smoother at $\widehat{\theta}_{k-1}$ to obtain $\nabla_{\theta} \log \widehat{p}_{\theta}( y_{1:T} )$.
		\STATE Apply the update $\widehat{\theta}_k = \widehat{\theta}_{k-1} + \gamma \cdot k^{-\alpha}\nabla_{\theta} \log \widehat{p}_{\theta}( y_{1:T} )$.
		\STATE Set $k = k + 1$.
	\ENDWHILE
	\STATE Set $\widehat{\theta} = \widehat{\theta}_k$.
\end{algorithmic}
\end{footnotesize}
\label{alg:doNLSSM}
\end{algorithm}	

\paragraph{Direct optimization -- particle based gradient ascent}
For this implementation, we make use of the approach from \citet{PoyiadjisDS:2011}, which is summarised in Algorithm~\ref{alg:doNLSSM}. To improve the numerical performance, the inference is carried out over the transformed parameters $\tilde{\phi} = \tanh^{-1}(\phi)$ and $\tilde{\tau} = \log (\tau)$. Hence, the two parameters are now unconstrained and these types of transformations can often result in beneficial variance reduction.

The gradient of the log-likelihood $\nabla_\theta \log p_\theta(y_{1:T})_{|\theta=\theta_{k-1}}$ is estimated by the Fisher identity using the fast forward-filtering backward-smoother (FFBSi) with early stopping as discussed by \cite{TaghaviLSS:2013}. We make use of $500$ forward particles, $100$ backward trajectories and rejection sampling for $75$ trajectories. For the Fisher identity, we require calculating the gradients of the complete data log-likelihood (with respect to the transformed parameters). Note first that the complete data log-likelihood is given by
\begin{align}
  \nonumber
   \log {}&p_{\theta}( x_{1:T}, y_{1:T} ) =  \log p_{\theta}( x_1 ) + \sum_{t=1}^{T-1} \log p_\theta(x_{t+1}\mid x_{t}) + \text{const.} \\
   \label{eq:app:nl:complete_data_log_lik}
   &= \frac{1}{2} \left\{ \log((1-\phi^2)\tau) - (1-\phi^2)\tau x_1^2
     + \sum_{t=1}^{T-1} \left( \log \tau - \tau (x_{t+1}-\phi x_t)^2 \right) \right\} + \text{const.}
\end{align}
We thus get,
\begin{align*}
  \frac{\partial}{\partial \tilde \phi}  &\log {}p_{\theta}( x_{1:T}, y_{1:T} )
  = \frac{\partial}{\partial \phi} \left\{   \log {}p_{\theta}( x_{1:T}, y_{1:T} ) \right\} \left( \frac{d\tilde\phi}{d\phi} \right)^{-1} \\
&= -\phi + (1-\phi^2)\tau \left\{ x_1^2 + \sum_{t=1}^{T-1} x_t (x_{t+1} - \phi x_t) \right\},
\end{align*}
where we have used the fact that $\frac{d}{d\phi}\tanh^{-1}(\phi) = (1-\phi^2)^{-1}$. Furthermore,
we have,
\begin{align*}
  \frac{\partial}{\partial \tilde \tau}  &\log {}p_{\theta}( x_{1:T}, y_{1:T} )
  = \frac{\partial}{\partial \tau} \left\{   \log {}p_{\theta}( x_{1:T}, y_{1:T} ) \right\} \left( \frac{d\tilde\tau}{d\tau} \right)^{-1} \\
&= \frac{1}{2} \left\{ T - \tau (1-\phi^2)x_1^2 - \tau \sum_{t=1}^{T-1} (x_{t+1}-\phi x_t)^2 \right\}.
\end{align*}
The optimisation is initialised in (untransformed) parameters $\{\phi,\tau\}=\{0.95, 10\}$ with $\alpha = -2/3$, $\gamma=0.01$ and runs for $K=250$ iterations.

\paragraph{Metropolis Hastings -- PMH}

%

The sampler is implemented in two steps. In the first step the smooth particle filter \citep{MalikPitt2011} is used with $500$ particles to get an initialisation of the parameters and to estimate the Hessian of the log-likelihood. The optimisation of the log-likelihood is done using a bounded limited-memory BFGS optimizer and the Hessian is estimated numerically using a central finite difference scheme. The resulting estimates of the parameters and inverse Hessian are
\begin{align*}
	\widehat{\theta}_{\text{ML}} = \{0.95, 0.02\}
	\qquad
	\widehat{\mathcal{I}}(\widehat{\theta}_{\text{ML}})
	=
	10^{-5}
	\begin{bmatrix}
	9.30 & 2.96 \\ 2.96 & 1.99
	\end{bmatrix}.
\end{align*}
The PMH0 algorithm is initialised in $\widehat{\theta}_{\text{ML}}$ and makes use of the bootstrap particle filter with $1 \thinspace 000$ particles to estimate the log-likelihood. The proposal is selected using the rules-of-thumb in \citet{SherlockThieryRobetsRosenthal2013} as
\begin{align*}
	q(\theta'' | \theta')
	=
	\mathcal{N}(\theta''; \theta', ( 2.562^2 / 2) \widehat{\mathcal{I}}(\widehat{\theta}_{\text{ML}})).
\end{align*}
We use $15 \thinspace 000$ iterations (discarding the first $2 \thinspace 000$ iterations as burn-in) to estimate the posteriors and their means.

\subsubsection{Data augmentation}

\paragraph{Expectation Maximisation -- PSAEM}
We outline the implementation details for the Particle SAEM algorithm (see Section~\ref{sec:EMrevisited}),
but the implementation for the PSEM algorithm (see Section~\ref{sec:DAEM}) follows similarly.

Note that particle SAEM requires us to compute an approximation of the $\Q$-function according to \eqref{eq:SAEM}.
The complete data log-likelihood can be written as (see \eqref{eq:app:nl:complete_data_log_lik}),
\begin{align*}
   \log {}&p_{\theta}( x_{1:T}, y_{1:T} ) \\
   &= \frac{1}{2} \left\{ \log((1-\phi^2)\tau) - (1-\phi^2)\tau x_1^2
     + \sum_{t=1}^{T-1} \left( \log \tau - \tau (x_{t+1}-\phi x_t)^2 \right) \right\} + \text{const.}
\end{align*}
If we define the complete data \emph{sufficient statistics} as $\mathcal{S} := (\Psi, \Phi, \Sigma, X)^\+ \in \mathbb{R}^4$ with
\begin{align*}
\Psi &= \frac{1}{T-1} \sum_{t=1}^{T-1} x_{t+1}x_{t}, &
\Phi &= \frac{1}{T-1} \sum_{t=2}^T x_{t}^2, &
\Sigma &= \frac{1}{T-1} \sum_{t=1}^{T-1} x_t^2, &
X &= x_1^2,
\end{align*}
we can thus write $\log p_{\theta}( x_{1:T}, y_{1:T} ) = -0.5 f(\theta ; \mathcal{S}) + \text{const}.$, where the function $f$ is defined as:
\begin{align}
  \nonumber
  f(\theta; \mathcal{S} ) := &-\log((1-\phi^2)\tau) + X(1-\phi^2)\tau \\
  \label{eq:app:nl:f-def}
  &+ (T-1) \left\{ -\log \tau + \tau (\Phi - 2\Psi\phi + \phi^2\Sigma) \right\}.
\end{align}
Expressing the complete data log-likelihood in terms of its sufficient statistics in this way is useful, since
it allows us to write the approximation of the $\Q$-function in \eqref{eq:SAEM} as:
\begin{align*}
  \widehat{\Q}_{k}(\theta) = -0.5 f(\theta ; \widehat{\mathcal{S}}_k) + \text{const.},
\end{align*}
where $\widehat{\mathcal{S}}_k = (\widehat \Psi_k, \widehat\Phi_k, \widehat\Sigma_k, \widehat X_k)^\+$ is a stochastic approximation of the sufficient statistics, computed recursively as
\begin{align*}
  \widehat \Psi_k &= (1-\alpha_{k}) \widehat \Psi_{k-1}  + \frac{\alpha_k}{T-1} \sum_{t=1}^{T-1} \left( \sum_{i=1}^N w_T^i[k] x_{t+1}^i[k] x_t^i[k] \right), \\
  \widehat \Phi_k &= (1-\alpha_{k}) \widehat \Phi_{k-1} + \frac{\alpha_k}{T-1} \sum_{t=2}^{T} \left( \sum_{i=1}^N w_T^i[k] (x_{t}^i[k])^2 \right), \\
  \widehat \Sigma_k &= (1-\alpha_{k}) \widehat \Sigma_{k-1}+ \frac{\alpha_k}{T-1} \sum_{t=1}^{T-1} \left( \sum_{i=1}^N w_T^i[k] (x_{t}^i[k])^2 \right), \\
  \widehat X_k &= (1-\alpha_{k}) \widehat X_{k-1} + \alpha_k \sum_{i=1}^N w_T^i[k] (x_{1}^i[k])^2,
\end{align*}
where $\{x_{1:T}^i[k], w_T^i[k]\}_{i=1}^N$ are the particle trajectories generated by the PGAS algorithm at iteration $k$.

Maximising $\widehat{\Q}_{k}(\theta)$ in the M-step of the algorithm is thus equivalent to minimising $f(\theta ; \widehat{\mathcal{S}}_k)$.
Let us therefore turn to the problem of minimising $f(\theta ; \mathcal{S})$ for an arbitrary (but fixed) value of the sufficient statistics $\mathcal{S}$.
First, noting that the leading two terms in \eqref{eq:app:nl:f-def} originate from the initial condition, which should have a negligible effect on the maximising argument for large $T$,
a good initialisation for the maximisation can be obtained by approximating
\begin{align*}
    f(\theta; \mathcal{S} ) \approx (T-1) \left\{ -\log \tau + \tau (\Phi - 2\Psi\phi + \phi^2\Sigma) \right\}.
\end{align*}
Indeed, minimising this approximation can be done on closed form, suggesting that
\begin{align*}
  \phi_{\text{opt.}} &\approx \Psi/\Sigma \\
  \tau_{\text{opt.}} &\approx (\Phi-\Psi^2/\Sigma)^{-1}.
\end{align*}
This provides us with a good initialisation for a numerical optimisation method which can be used to minimise
$ f(\theta; \mathcal{S} )$ to desired precision.

\paragraph{PGAS} 
In the PGAS algorithm for Bayesian inference we employ a Gibbs sampler, iteratively simulating $x_{1:T}$ from the PGAS
kernel, and $\theta$ from the conditional distribution $p(\theta \mid x_{1:T}, y_{1:T})$.
This distribution is given by
\begin{align}
  p(\phi, \tau | x_{1:T}, y_{1:T}) = \frac{1}{Z} \tau^{a +{T \over 2}- 1} e^{-\tilde b \tau} \mathbbm{1}_{\{ |\phi| \leq 1 \}} \sqrt{1-\phi^2} e^{ -\frac{\tilde \tau}{2} \left( \phi - \tilde \mu \right)^2 },
\label{eq:app:pgas:cond}
\end{align}
where the constants are given as follows:
\begin{align}
\tilde b &= b + {1 \over 2} \sum_{t=1}^T x_t^2 - {1 \over 2} \frac{\left( \sum_{t=1}^{T-1} x_{t+1} x_t \right)^2}{\sum_{t=2}^{T-1} x_t^2}, \\
\tilde \tau &= \tau \sum_{t=2}^{T-1} x_t^2, \\
\tilde \mu &= \frac{\sum_{t=1}^{T-1} x_{t+1} x_t}{\sum_{t=2}^{T-1} x_t^2}.
\end{align}
Simulating from \eqref{eq:app:pgas:cond} is done by rejection sampling with an instrumental distribution,
\begin{equation}
q(\phi, \tau | x_{1:T}, y_{1:T}) = \mathcal{G} \left( \tau; a + {T-1 \over 2}, \tilde b \right) \N \left( \phi; \tilde \mu, {\tilde \tau}^{-1} \right).
\end{equation}%
Specifically, we propose a draw $(\phi', \tau')$ from the instrumental distribution and accept this as a draw from \eqref{eq:app:pgas:cond}
with probability ${\mathbbm{1}_{\{ |\phi| \leq 1 \}} \sqrt{1-\phi^2}}$.


\clearpage
\bibliographystyle{plainnat}
\bibliography{PF4SI}
\end{document}